\documentclass[AMA,STIX1COL]{WileyNJD-v2}

\DeclareMathOperator*{\argmin}{argmin}

\articletype{Article Type}%

\usepackage{colortbl}

\received{26 April 2016}
\revised{6 June 2016}
\accepted{6 June 2016}

\newcommand*{\red}{\textcolor{red}}

\begin{document}

\title{Balancing events, not patients, maximizes power of the logrank test: and other insights on unequal randomization in survival trials}

\author[1]{Godwin Yung*}

\author[2]{Kaspar Rufibach}

\author[2]{Marcel Wolbers}

\author[1]{Ray Lin}

\author[3]{Yi Liu}

\authormark{AUTHOR ONE \textsc{et al}}

\address[1]{\orgdiv{Statistical Methods Collaboration and Outreach Group}, \orgname{F. Hoffmann-La Roche AG}, \orgaddress{\state{California}, \country{USA}}}

\address[2]{\orgdiv{Statistical Methods Collaboration and Outreach Group}, \orgname{F. Hoffmann-La Roche AG}, \orgaddress{\state{Basel}, \country{Switzerland}}}

\address[3]{\orgdiv{Department of Biostatistics}, \orgname{Nektar Therapeutics}, \orgaddress{\state{California}, \country{USA}}}

\corres{*Godwin Yung, Genentech, 1 DNA Way, South San Francisco, CA 94080. \email{yungg@gene.com}}


\abstract[Summary]{We revisit the question of what randomization ratio (RR) maximizes power of the logrank test (LRT) in event-driven survival trials under proportional hazards (PH). By comparing three approximations of the LRT (Schoenfeld, Freedman, and Rubinstein) to empirical simulations, we find that the RR that maximizes power is the RR that balances number of events across treatment arms at the end of the trial. This contradicts the common misconception implied by Schoenfeld's approximation that 1:1 randomization maximizes power. Besides power, we consider other factors that might influence the choice of RR (accrual, trial duration, sample size, etc.). We perform simulations to better understand how unequal randomization might impact these factors in practice. Altogether, we derive 5 insights to guide statisticians in the design of survival trials considering unequal randomization.}



\maketitle


\section{Introduction}\label{sec:intro}
It is common practice in pivotal survival trials to
\begin{enumerate}
    \item randomize patients 1:1 to an experimental and control arm, and to
    \item follow patients until a certain total number of events have been observed.
\end{enumerate}
Practice 2 can be attributed to Schoenfeld\cite{Schoenfeld1981}, whose approximation for the large-sample distribution of the logrank test (LRT) established number of events as the effective sample size under proportional hazards (PH). Practice 1, on the other hand, may involve multiple considerations, including patient preference, cost, ethics, and statistical power.\cite{Pocock1979, Avins1998, Dumville2006, Hey2014, Peckham2015} With respect to power, it is often believed that 1:1 randomization maximizes power of the LRT under PH. From the authors' collective experience, this belief has been perpetuated by academic, industry, and government statisticians alike. This belief is however unfounded. Randomizing patients 1:1 does not in fact maximize power in event-driven trials under PH.

This paper seeks to rectify the common misconception that 1:1 randomization maximizes power of the LRT in event-driven survival trials under PH. To that end, it is helpful to consider why this misconception exists to begin with. Perhaps it is because 1:1 randomization maximizes power in other settings, e.g., continuous endpoint with common variance. Or because Schoenfeld's approximation for the LRT implies that 1:1 maximizes power under PH. We shall later show that Schoenfeld's approximation increasingly underestimates power when more patients are randomized to the experimental arm than the control arm. Of course, an equation that is biased in favor of 1:1 cannot itself be used to justify that 1:1 randomization maximizes power.

Besides Schoenfeld, other approximations for the LRT under PH have been proposed which also have immediate, albeit different, implications for randomization. Sposto used Rubinstein's approximation for the LRT to suggest that power can be gained by randomizing more patients to the experimental arm.\cite{Rubinstein1981, Sposto1987} Hsieh used Freedman's approximation to suggest that power is maximized when the randomization ratio is equal to 1/(hazard ratio).\cite{Freedman1982, Hsieh1992}

But despite these previous publications, a question remains unclear. What randomization ratio (RR) maximizes power in event-driven trials if it isn't 1:1? Sposto considered time-driven trials (i.e., trials in which patients are followed for a fixed period of time), which is different from common practice (Practice 2). Hsieh only compared Freedman's approximation to Schoenfeld's. Freedman's approximation could thus face the same challenge as Schoenfeld's. That is, if it is found to have differential accuracy under unequal randomization, then its conclusion of when power is maximized is equally questionable. \red{Abel, Jensen, et al.\cite{Abel2015} identified seven issues related to the assumptions, applicability, and practical use of Schoenfeld and Freedman's approximations. However, randomization ratio was not one of the topics explored by the authors.}

The rest of the article is organized as follows. In Section \ref{sec:logrank}, we review Schoenfeld's, Freedman's, and Rubinstein's approximations for the LRT and their implications on randomization in event-driven trials under PH. We then compare their power calculations to empirical simulations over a wide range of scenarios. Our simulation study will reveal that Rubinstein's approximation is the most accurate of the three, and that it consistently estimates empirical power within 1\% absolute error. This will lead us to the conclusion that, for event-driven trials under PH, the RR that maximizes power of the LRT is the RR that balances number of events across treatment arms at the end of the trial. \red{A helpful way to remember this is to remember to balance the \emph{effective sample size} across treatment arms. For continuous endpoints, that means balancing number of patients. For time-to-event endpoints, that means balancing number of events.}

Power is just one of many considerations when it comes to actually selecting the RR in a clinical trial. In Section \ref{sec:factors}, we review some of these other considerations, including accrual duration, trial duration, and sample size---design characteristics that are important for trial statisticians to consider, but that have been discussed in less detail previously.

In Section \ref{sec:case}, we compare multiple trial designs with the same power but different randomization ratio, accrual duration, trial duration, and/or sample size. The objective here is to generate insights into when certain trial designs may be attractive in practice. We will find that 3:2 randomization generally leads to a minor delay in trial readout if accrual rate and sample size do not change. However, delay may be avoided by a moderate increase in accrual rate or minor increase in number of patients. 2:1 randomization, on the other hand, can result in a significant delay in trial readout or increase in sample size, thereby delaying access to medicine or increasing the financial burden of clinical trials.

While the focus of our paper is on survival trials under PH and with a single readout, we recognize that there are trials in which one would expect non-proportional hazards (NPH) at the time of trial design, e.g., trial comparing immunotherapy vs. chemotherapy. It is also common for Ph3 oncology trials to conduct interim analyses with the hope of bringing medicines to patients earlier. In Section \ref{sec:discuss}, we conclude with a brief discussion of these two topics and a summary list of our learnings.

\section{Large-sample distribution of the logrank statistic}\label{sec:logrank}

Consider a survival trial with $n_e = n \pi$ patients randomized to the experimental arm and $n_c = n (1 - \pi)$ to the control, $0 < \pi < 1$. The randomization ratio is given by $\phi$:1 where $\phi=\pi / (1-\pi)$. Patients are enrolled over an accrual period of $r$ months and followed until a total of $d$ events have been observed. 
\red{We denote by $t_d$ the total trial duration and by $S_e(\cdot)$ and $S_c(\cdot)$ the event survival functions for the experimental and control arm, respectively.} The logrank test will serve as the primary analysis for assessing equality of survival, i.e., testing the null hypothesis $H_0$: $S_e(\cdot) \equiv S_c(\cdot)$ versus the alternative $H_1$: $S_e(\cdot) \nequiv S_c(\cdot)$.

To assess power of the LRT, one needs the distribution of the logrank statistic under the alternative hypothesis that treatment is effective. For that purpose, a number of large-sample approximations have been proposed under PH (i.e., hazard ratio is equal to some constant $e^{\theta}=\log S_e(t)/\log S_c(t)$ for all $t$) which take the form of a normal distribution with some mean $\mu$ and unit variance. Under such normal approximations, the $\phi$ which maximizes statistical power is equivalent to that which minimizes $\mu$. 

Next, we review several $\mu$'s and their implied $\argmin_\phi \mu$. Before doing so, it is worth noting that these approximations were originally proposed under the setting that patients are followed until a certain amount of time has elapsed. Thus, some $\mu$'s may in fact depend on trial duration. But as explained by Yung and Liu\cite{Yung2020}, these approximations can be applied to event-driven trials by approximating trial duration with $t_d = \inf \{t: E(D(t))=d\}$, the earliest time at which the expected number of accumulated events $E(D(t))$ equals $d$.

\subsection{Schoenfeld (1981)}
Under local alternatives (i.e., an alternative that converges to $H_0$ as $n$ increases) \red{and assuming that censoring due to loss of follow-up is independent of survival and equal across treatment arms}, Schoenfeld\cite{Schoenfeld1981} approximated the large-sample mean of the logrank statistic by
\begin{equation}
    \mu_{S} = \frac{\theta}{1+\phi} \sqrt{d \phi}.
\end{equation}
\red{This result is somewhat contradicting, given that two different treatment effects (local alternatives and fixed treatment effect $\theta$) were assumed to approximate/simplify different parts of LRT's asymptotic distribution.} It is, however, also elegant. $\mu_S$ depends on $S_e(\cdot)$ and $S_c(\cdot)$ only through the log hazard ratio $\theta$. It does not change regardless of the censoring or accrual distributions. The fact that $\mu_S$ depends on $d$, but not $n$ or $t_d$, gives investigators the flexibility to design event-driven trials that are fast (large $n$, small $t_d$) or slow (small $n$, large $t_d$) depending on the patient population and available resources.\cite{Sandoval2021}   

It can be shown by elementary calculus that
\begin{equation}
\label{eq:schoenfeld}
    \argmin_{\phi} \mu_S(\phi;d, \theta) = 1.
\end{equation}
Thus, for event-driven trials under PH, $\mu_S$ implies that power is maximized by randomizing patients 1:1.

\subsection{Freedman (1982)}
Freedman\cite{Freedman1982} approximated the large-sample mean of the logrank statistic by
\begin{equation}
    \mu_{F} = \frac{e^{\theta}-1}{1+e^{\theta}\phi} \sqrt{d \phi}.
\end{equation}
Their derivation conditions on the set of patients at risk before each event and assumes that the ratio of patients at risk in the treatment and control arm is always equal to 1. Like $\mu_S$, $\mu_F$ depends only on $\phi$, $d$ and $\theta$. Hsieh\cite{Hsieh1992} compared $\mu_F$ to $\mu_S$ and concluded that $\mu_F$ is more accurate for values of $\phi$ away from 1, but less accurate for values of $\phi$ close to 1. Related to our discussion, the authors also noted that
\begin{equation}
\label{eq:freedman}
    \argmin_{\phi} \mu_F(\phi;d, \theta) = e^{-\theta}.
\end{equation}
For example, under hazard ratios 2/3 and 1/2, randomizing patients 3:2 and 2:1 maximizes power, respectively.

\subsection{Rubinstein (1981)}
\red{Suppose events in the experimental and control arms follow exponential distributions with hazard rates $\lambda_e$ and $\lambda_c$, respectively. Under arbitrary accrual distribution and arbitrary censoring distributions that are independent of survival but not necessarily equal across treatment arms}, Rubinstein\cite{Rubinstein1981} derived the maximum likelihood estimates (MLEs) for the hazard rates and the corresponding test for $H_0$: $\log(\lambda_e / \lambda_c)=0$. Their wald test based on the exponential MLEs can be used to approximate the LRT under the argument that the LRT is asymptotically efficient when there is general noninformative censoring.\cite{Peto1972} Doing so, we obtain
\begin{equation}
    \mu_{R} = \theta \left (\frac{1}{E(D_e(t_d))} + \frac{1}{E(D_c(t_d))} \right )^{-1/2},
\end{equation}
where $D_e(t)$ and $D_c(t)$ denote the accumulated number of events in the experimental and control arm by calendar time $t$. Since $E(D_e(t_d)) + E(D_c(t_d)) = d$ in event-driven trials, it follows that
\begin{equation}
\label{eq:rubinstein}
    \argmin_{\phi} \mu_R(\phi;d, \theta, A, S_c, L) = \left \{\phi \mbox{ such that } E(D_e(t_d)) = E(D_c(t))) \right \}.
\end{equation}
Rubinstein's approximation therefore implies that power is maximized when the expected number of events at the end of the trial is equal across treatment arms. 

To gain some intuition as to how this conclusion compares to Schoenfeld's and Freedman's, suppose that accrual follows a uniform distribution and that dropout in both arms follows an exponential distribution with rate $\eta$. We show in Appendix \ref{app1} that
\begin{equation}
    \lim_{t \rightarrow 0^+} \frac{E(D_e(t))}{E(D_c(t))} =\phi e^{\theta}
\end{equation}
and
\begin{equation}
    \lim_{t \rightarrow \infty} \frac{E(D_e(t))}{E(D_c(t))} =\phi \frac{\lambda_e / (\lambda_e + \eta)}{\lambda_c / (\lambda_c + \eta)}.
\end{equation}
Thus, for extremely short trials (i.e., $t_d \approx 0$), the ratio of events contributed by each arm is approximately $\phi e^{\theta}$ and balance of events at the end of trial can be achieved by setting $\phi = e^{-\theta}$. Meanwhile for long trials with almost every patient observing an event (i.e., $E(D(t_d)) \approx E(D(\infty))$ and $\eta \approx 0$), the ratio of events contributed by each arm is approximately $\phi$ and balance of events at the end of trial can be achieved by setting $\phi = 1$. Taken together, one may consider Equation (\ref{eq:rubinstein}) as lying somewhere  between Equations (\ref{eq:schoenfeld}) and (\ref{eq:freedman}) depending on the trial duration and dropout rate.

\subsection{Which approximation is the most accurate?}
\label{sec:accuracy}

We now consider which of the three approximations---Schoenfeld, Freedman, or Rubinstein---is the most accurate for estimating power and for comparing power between different randomization ratios. To that end, suppose that survival in the control arm follows an exponential distribution with median of 12 months and that dropout follows an exponential distribution such that patients have a 1\% probability of dropping out every 12 months. We consider event-driven trials with varying treatment effect and trial duration by defining a grid of values for the hazard ratio (HR) and event-patient ratio (EPR $=d/n$):  $(\text{HR}, \text{EPR}) \in [0.5, 0.8] \times [0.5, 0.8]$. For each unique pair (HR, EPR), we set
\begin{itemize}
    \item number of events $d = \left\lceil (1.96+0.84)^2 / (0.5 \times 0.5 \times (\log \text{HR})^2) \right\rceil$ based on Schoenfeld's approximation for 1:1 randomization, one-sided $\alpha=0.025$ and 80\% power,
    \item sample size $n = d \times (d/n)^{-1}$, and
    \item accrual rate $20+(50-20)(\text{HR}-0.5)/(0.8-0.5) $, which increases from 20 to 50 patients/month as HR increases so that the expected total trial duration is always less than 5 years.
\end{itemize}
We then calculate power $\text{Power}(\text{RR}, \text{M};\text{HR}, \text{EPR})$ for different randomization ratios RR (1:1, 3:2, 2:1) using different methods M (S = Schoenfeld, F = Freedman, R = Rubinstein, E = Empirical simulations).

\begin{figure}[t]
    \centering
    \includegraphics[width=0.8\textwidth]{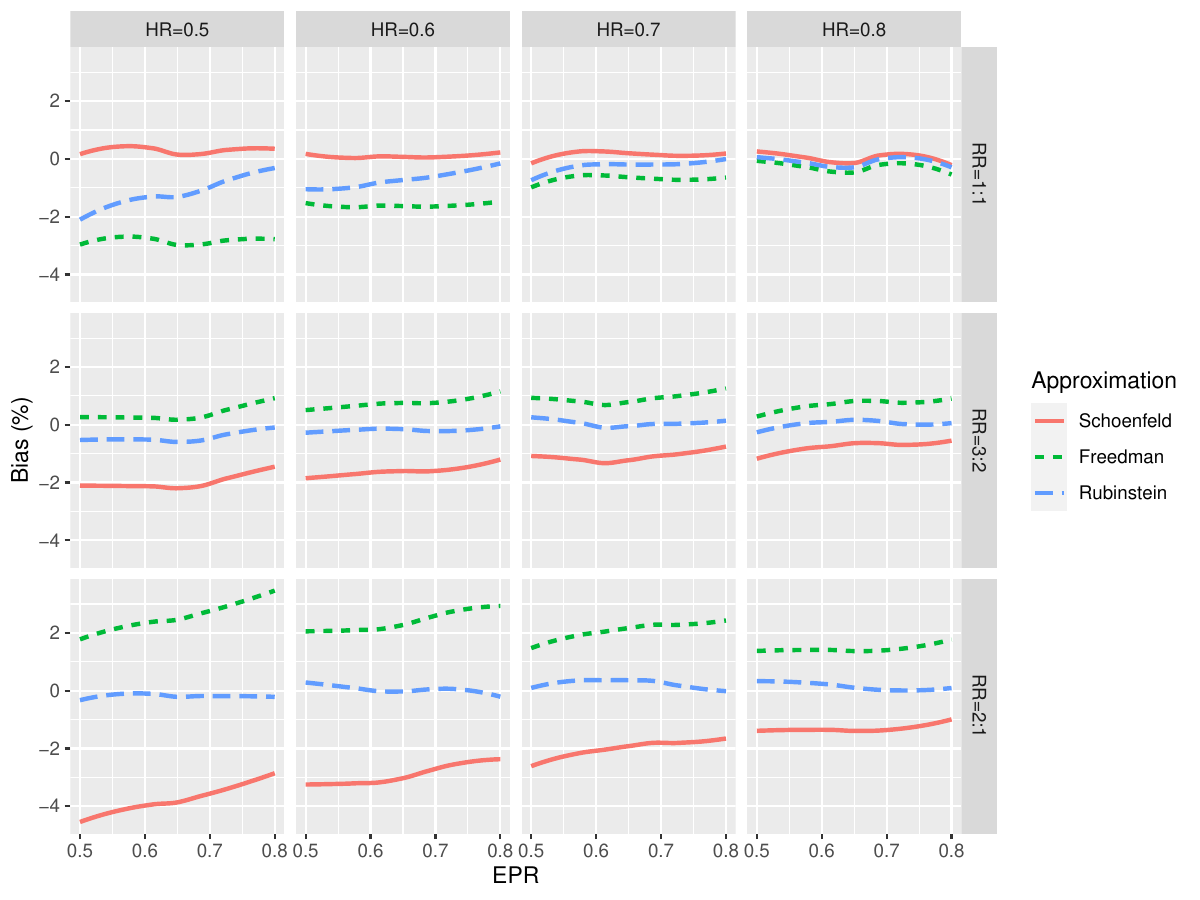}
    \caption{Bias of power calculations comparing various approximations to empirical simulations across scenarios with different event-patient ratios (EPR), hazard ratios (HR) and randomization ratios (RR). LOESS smoothing was applied to more clearly display trends amidst Monte Carlo error. The Monte Carlo standard error is estimated to be $\sim$0.6\% given 5000 simulations and true power $\sim$80\%. Note that remaining minor fluctuations in the curves is to be expected, since the x-axis may not exactly represent the actual EPR for a specific scenario; actual EPR may differ slightly from the desired EPR (x-axis) since it also depends on the RR.}
    \label{fig:accuracy1}
\end{figure}

Figure \ref{fig:accuracy1} shows the bias of power calculations for the different approximations compared to 5000 empirical simulations:
\begin{equation}
    \text{Bias} = \text{Power}(\text{RR}, \text{M}; \text{HR}, \text{EPR})-\text{Power}(\text{RR}, \text{E}; \text{HR}, \text{EPR}).
    \label{eq:bias}
\end{equation}
As noted by Barthel et al\cite{Barthel2006}, Schoenfeld performs well under 1:1 even for HRs as small as 0.5. Although the authors did not consider performance under unequal randomization, here we see that Schoenfeld increasingly underestimates empirical power under 3:2 and 2:1 randomization, particularly for effective treatments and fast trials (small HR and EPR). For example, when RR = 2:1 and HR = EPR = 0.5, Schoenfeld underestimates power by more than 4\%. Freedman may also lead to over- or under-estimation by up to 3\% under certain scenarios. Only Rubinstein consistently estimates empirical power accurately within 1\%.  

\begin{figure}[t]
    \centering
    \includegraphics[width=0.8\textwidth]{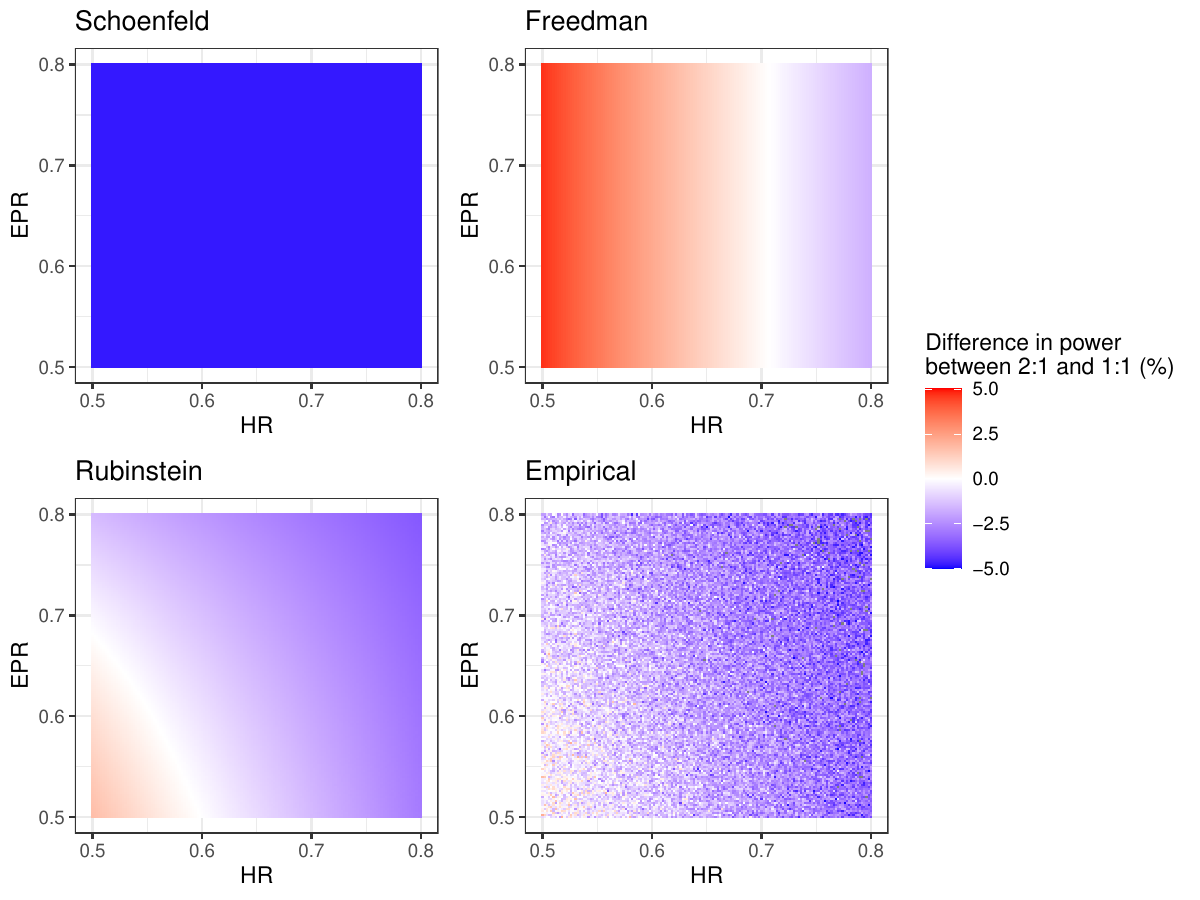}
    \caption{Difference in power between 2:1 and 1:1 randomization, according to various approximations or empirical simulations and across scenarios with different hazard ratios (HR) and event-patient ratios (EPR). For difference in power between 3:2 and 1:1 randomization, see Figure S1.}
    \label{fig:accuracy2}
\end{figure}

Because Schoenfeld increasingly underestimates power under unequal randomization, one can imagine that it can lead to wrongful conclusions when comparing power between randomization ratios. Figure \ref{fig:accuracy2} shows the difference in power between 2:1 and 1:1 randomization according to the various methods of calculation, i.e.,
\begin{equation*}
    \text{Difference in power} = \text{Power}(\text{2:1}, \text{M}; \text{HR}, \text{EPR})-\text{Power}(\text{1:1}, \text{M}; \text{HR}, \text{EPR})
\end{equation*}
Schoenfeld suggests that randomizing patients 2:1 instead of 1:1 decreases power by 4.8\%, regardless of HR or EPR. Meanwhile, Freedman suggests that difference in power depends on the HR, with 2:1 randomization becoming increasingly favorable as HR decreases (e.g., when HR=0.5, 2:1 randomization increases power by 4.6\%). Rubinstein suggests that difference in power depends on both HR and EPR. Compared to Freedman, the potential for increase in power is dampened. Only when both HR and EPR are small (<0.6) does 2:1 randomization lead to a marginal increase in power (0-2\%). Similar observations can be made comparing 3:2 to 1:1 randomization (Figure S1).

\begin{figure}[t]
    \centering
    \includegraphics[width=0.8\textwidth]{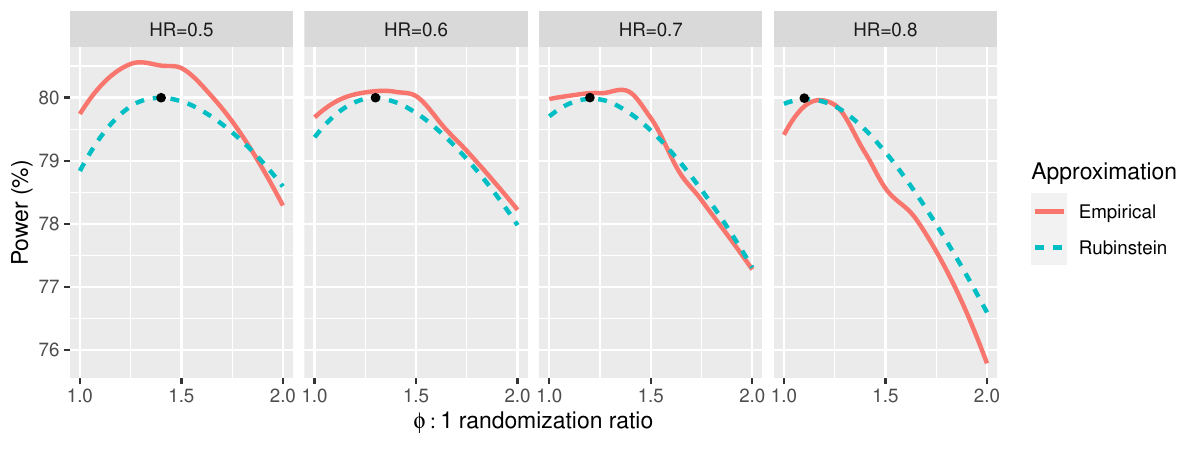}
    \caption{Power for various randomization ratios, according to 10,000 empirical simulations and Rubinstein's approximation. Four scenarios with varying hazard ratios (HR) and event-patient ratio $0.7$ are considered. Black dots indicate the point at which power is maximized under Rubinstein's approximation. They also indicate the randomization ratio under which events are expected to be balanced across treatment arms at the end of the trial. LOESS smoothing was applied to the empirical curves to more clearly display trends amidst Monte Carlo error. The Monte Carlo standard error is estimated to be $\sim$0.4\% given 10,000 simulations and true power $\sim$80\%.}
    \label{fig:optimal_rr}
\end{figure}

Having compared Schoenfeld's, Freedman's, and Rubinstein's approximations to empirical simulations across many scenarios, it is clear that Rubinstein's is the most accurate of the three. It consistently estimates power well (Figure \ref{fig:accuracy1}) and captures the impact of HR and EPR on power for unequal vs. equal randomization most accurately (Figure \ref{fig:accuracy2}). One only needs to look at specific scenarios to confirm that the randomization ratio that maximizes power is indeed the randomization ratio that balances number of events across treatment arms at the end of the trial---as suggested by Rubinstein's approximation. Figure \ref{fig:optimal_rr} illustrates this in four scenarios. (For more scenarios, see Figure S2.)

This conclusion that ``power is maximized when events are balanced across arms'' is not completely new. Hsieh\cite{Hsieh1992} hypothesized that ``the optimal design may exist in the neighborhood from equal allocation of sample sizes to equal allocation of events''. Later, Yung and Liu\cite{Yung2020} narrowed it down to when ``events are balanced across arms''. However, Yung and Liu drew their conclusion from observing the application of another normal approximation for the logrank statistic. This approximation---while more flexible because it can be used under arbitrary survival distributions---cannot be rewritten in such a way as Rubinstein's to arrive at Equation (\ref{eq:rubinstein}). \red{Therefore, the current work can be seen as providing further substantiation of Yung and Liu’s observation.}


\red{While 1:1 may not maximize power, we see from Figure \ref{fig:optimal_rr} that its power is often close to the randomization ratio that does.} Only in the most extreme scenario (HR = EPR = 0.5) does 1:1 randomization lead to a 2\% loss of power. This observation is consistent with Sposto's conclusion for time-driven trials: ``Although one can compute the allocation which maximizes power, the resulting increase in power would be inconsequential. Rather, the fact that a range of allocations yields practically the same power allows one to consider unequal allocations when they would be desirable for other reasons''.\cite{Sposto1987} This ties nicely into Section \ref{sec:factors} where we consider some of these ``other reasons''.

\subsection{Generalizing Rubinstein's Conclusion}\label{sec:generalizing_rubinstein} 
\red{Rubinstein's approximation was derived under exponential survival, arbitrary accrual, and arbitrary censoring distributions. Its accuracy was further illustrated in the previous subsection under exponential survival.} Therefore, one might wonder what happens under general PH. Can we still conclude that power of the LRT is maximized by balancing the expected number of events across arms?

Wu\cite{Wu2015} considered Weibull survival distributions under PH and showed that the asymptotic mean of the parametric test statistic is equal to $\mu_R$ times the common shape parameter. Thus, our conclusion is at least robust to the Weibull distribution. 
\red{For the case under general PH, an argument proceeds as follows. Given two survival functions under PH, there exists a monotonic transformation of the survival times such that the transformed observations follow two exponential distributions.\cite{Fay2018} Rubinstein's approximation implies that power of the LRT---when applied to the transformed observations---is maximized when events are balanced across arms. Meanwhile, the LRT is a rank-based test, so a monotonic transformation does not change the log-rank statistic itself. The transformation also does not change the total number of events (assuming an event-driven trial) nor the ordering of events across arms. Therefore, if balancing events across arms maximizes power for the transformed observations (i.e., exponential survival curves), then it too maximizes power for the original observations (i.e, survival curves under general PH).}

\section{Considerations for selecting a randomization ratio}\label{sec:factors}

As shown in Section \ref{sec:logrank}, power of the LRT is maximized in event-driven survival trials under PH when the number of events is balanced across treatment arms. Since pivotal trials typically target a certain level of statistical power (e.g., 80\% or 90\%), a RR that is sub-optimal with respect to power may require more events to be observed, which may in turn impact trial duration or sample size. We now describe in more detail these design aspects along with other considerations when selecting a RR.

\subsection{Accrual and trial duration}
Investigators often cite accelerating patient recruitment as one of the reasons for implementing unequal randomization.\cite{Hey2014, Peckham2015} The assumption is that patients are more likely to participate in a randomized trial when their odds of receiving the experimental therapy are greater. This may be especially true if the control therapy is unattractive, such as a placebo. 

By shortening accrual duration, the implicit hope is that the overall trial duration will also be shortened so that scientific conclusions may be drawn earlier, patients may receive effective therapies sooner, and trial costs may be lowered. However, if the experimental therapy is superior to the control therapy, then assigning more patients to the experimental arm (and less patients to the control arm) will mean that longer follow-up is needed to observe the same number of events. Given that a trial is made up of an accrual and follow-up period, whether unequal randomization shortens or prolongs the total trial duration therefore depends on whether accelerated accrual offsets prolonged follow-up. 

In addition, it is unclear whether randomizing more patients to the experimental arm actually accelerates recruitment.\cite{Hey2014} The reality is that RR is just one of many factors that patients consider when deciding whether or not to participate in a trial. Other important factors include duration of treatment, mode of administration, out-of-pocket costs, travel time, and home visits.\cite{Eek2016} From a systems perspective, insufficient infrastructure and restrictive eligibility criteria also present significant barriers to clinical trial participation.\cite{Nipp2019, Wong2020}

\subsection{Sample size}
Sponsors of survival trials have flexibility when it comes to deciding the overall sample size of a trial. According to Schoenfeld's approximation, a trial under PH can achieve the target level of power under various number of patients, so long as the target number of events is observed. While sponsors can theoretically calculate number of patients with the intention of achieving balance of events at the time of analysis (thereby making the most efficient use of events), the standard practice is to target an EPR that balances resource constraints such as the available pool of patients, trial costs, and trial timeline.

Therefore, a more realistic consideration is as follows: If employing unequal randomization would increase trial duration (compared to equal randomization and assuming fixed accrual rate and sample size), how many additional patients would need to be recruited to avoid any delay in trial readout? Doing so would impact recruitment and trial costs, but it would also even the playing field with respect to trial duration. We shall explore ``edge cases'' such as this in Section \ref{sec:case}.

\subsection{Balance in baseline characteristics}
A related consideration to sample size is the balance in baseline characteristics. Unequal randomization is more likely to result in an imbalance in baseline characteristics between treatment arms and thus chance bias.\cite{Roberts1999} This concern applies more to small- to medium-sized trials as well as subgroups within a trial, since they carry a higher risk of imbalance to begin with. \red{Of course, if there are known factors that influence prognosis, then imbalance with respect to these prognostic factors can be prevented via stratified randomization.}

To illustrate the potential impact of unequal randomization, let $X$ be a baseline covariate with standard deviation $\sigma$. A trial with $n_{\pi}$ patients that randomizes $\pi\times100\%$ patients to the experimental arm and $(1-\pi)\times100\%$ patients to the control arm will have the following variance for the difference in average baseline:
\begin{equation*}
    \mbox{Var}(\bar X_1 - \bar X_0) = \frac{\sigma^2}{n_{\pi}} \cdot \frac{1}{\pi (1-\pi)}.    
\end{equation*}
This quantity is directly associated to the risk of baseline imbalance\red{, since the larger the variance, the more likely it is for the trial to observe a large difference in average baseline.} 

This quantity is also equivalent to the variance for the difference in average baseline in a trial employing 1:1 randomization, but with sample size
\begin{equation*}
    n_{0.5} = n_{\pi} \cdot \frac {\pi(1-\pi)}{0.5(1-0.5)}.
\end{equation*}
Thus, for example, a similar amount of baseline imbalance can be expected in a 2:1 randomized trial with 180 patients as in a 1:1 randomized trial with 160 patients.

\subsection{Secondary estimands}
Clinical trials often have multiple trial objectives. While stakeholders would like to see a primary analysis that is statistically significant and clinically relevant, they may also want to see ``sufficiently mature'' data on secondary objectives, so that they can make an informed judgement of the drug's overall benefit-risk profile. Sponsors would do well to evaluate the potential impact of RR on important secondary objectives and chosen measures of data maturity.

The ICH E9(R1) Addendum introduced the estimands framework as a structured approach to formulating clinical trial objectives.\cite{ICH2019} Under this framework, an estimand (i.e., a precise description of the treatment effect reflecting the clinical question posed by a clinical trial objective) consists of five attributes: population, treatment, endpoint, intercurrent events and strategy, and population-level summary. Below are a few examples of estimand attributes that may define important secondary objectives in oncology trials. These examples should not be considered as collectively defining a single estimand, but rather as individual, defining attributes that differentiate a secondary objective from the primary objective.
\begin{itemize}
    \item Treatment: experimental arm only (e.g., in a trial where the control arm is already well understood)
    \item Population: biomarker subgroup (e.g., for immune checkpoint inhibitors)
    \item Endpoint: overall survival (e.g., when primary endpoint is progression free survival)
    \item Intercurrent events and strategy: hypothetical strategy for subsequent therapies (e.g., trials where treatment switching to the active arm occurs)
    \item Summary measure: survival probability at 2 years
\end{itemize}

For any given secondary objective and chosen estimator, what constitutes as sufficient or mature follow-up needs to be defined.\cite{Clark2002, Pocock2002, Betensky2015, Rufibach2023} For example, one might consider a Kaplan-Meier estimate of the 2-year survival probability to be mature provided that the width of its confidence interval is within a certain range (``precision'' and ``reliability''), or that the estimate will change little given longer follow-up (``stability''). In other cases, defining mature follow-up may be difficult (e.g., overall survival in indolent cancers\cite{Fleming2024}).

To further illustrate how a study team might act taking into consideration one of the secondary objectives above, Hey\cite{Hey2014} suggested assigning more patients to the experimental arm if the control arm is well understood, so that more efficacy and safety information on the experimental therapy can be obtained. We have also heard hesitancy from fellow statisticians to employ unequal randomization for fear of key subgroups having imbalanced baseline characteristics or fewer events.

\subsection{Other}
If a treatment arm is substantially more expensive than the other, then it may be desirable to assign more patients to the less expensive intervention treatment.\cite{Torgerson2000, Chandereng2020} For example, CAR T-cell therapies carry high costs because of their collection and manufacturing process. Meanwhile, therapies that were recently approved and are currently under exclusivity may be expensive for sponsors of a new trial to obtain as the control therapy.

Some have also argued that pivotal trials start from a position of clinical equipoise, and that it is incumbent upon the sponsors to demonstrate efficacy and safety. Therefore, randomizing more patients to the experimental arm at the start of a trial---when a favorable benefit-risk profile has yet to be established---is unethical.\cite{Hey2014} Others have argued the opposite, that pivotal trials rarely begin under true equipoise. This might be because a pivotal trial is initiated after promising efficacy and safety results were observed in an earlier phase study---in which case more patients should be randomized to the experimental arm.\cite{Avins1998} Or it might be because safety has already been well-established for the control therapy, but is of particular concern for the experimental therapy---in which case more patients should be randomized to the control arm. 

Randomizing more patients to experimental therapy may benefit enrolled patients by prolonging their time-to-event (improved individual ethics). However, it may also prolong trial duration, in which case patients not enrolled in the trial would have to wait longer for drug approval and access (worsened collective ethics). Other factors that may tip the argument in favor of individual or collective ethics include disease prevalence, disease severity, strength of early phase results, and safety profile of the experimental therapy.\cite{Senn2002}

\section{Case studies}\label{sec:case}
We now consider various trial designs employing 1:1 randomization and alternative designs that differ in randomization ratio, accrual duration, trial duration, and/or sample size. Per usual in trial design, power is fixed at some level, in this case 80\%. 

For the sake of clarity, we consider first a real trial design, that of Checkmate-017. Since there are infinitely many alternative designs to Checkmate-017, we consider six specific ones, ``edge cases'' that indicate the point at which a set of designs is more or less favorable than the original. This will allow us to limit our investigation without compromising the ability to generate comprehensive insight.

\subsection{Checkmate-017}\label{sec:cm017}
Checkmate-017 was a randomized, open-label phase 3 trial that compared nivolumab vs. docetaxel in patients with advanced squamous-cell non–small-cell lung cancer.\cite{Brahmer2015} The trial planned to enroll 264 patients over a period of 12 months, to randomize the patients 1:1 to nivolumab and docetaxel, and to follow for overall survival until 189 deaths were observed. Given two-sided alpha 0.04 (0.01 was reserved for the second primary endpoint overall response rate) and assuming exponential survival distributions with medians 11.4 and 7.0 months, the trial was projected to take 24 months to complete and to have 90\% statistical power to detect a target hazard ratio of $7.0/11.4=0.61$. 

For the purpose of this case-study, we assume that Checkmate-017's protocol follows similar assumptions as above (1:1 randomization, exponential survival distributions with medians 11.4 and 7.0 months, 22 patients/month enrollment rate, 0.72 event-patient ratio), but with the following differences and additional assumptions: trial is powered at 80\% given two-sided alpha 0.05, enrollment follows a uniform distribution, and dropout follows an exponential distribution so that patients have a 5\% probability of dropping out every 12 months. Consequently, this trial design plans to enroll 186 patients over a period of $\sim$8.5 months and to follow the patients until 133 deaths are observed for a total trial duration of $\sim$21.7 months (CM-017 in Table \ref{tab:checkmate}). 

\begin{center}
\begin{table*}[t]
    \centering
    \caption{Checkmate-017 protocol (CM-017) and six alternative trial designs (Alt) which are also powered at 80\%, but that differ by randomization ratio (\emph{RR}) and other design parameters as summarized by \emph{Description}. Number of events ($d$), number of patients ($n$), accrual duration, and expected trial duration for the alternative designs are either fixed to be similar to Checkmate-017 (indicated by parentheses) or estimated empirically via 10,000 simulations.}
    \label{tab:checkmate}
    \begin{tabular}{clcccc}
    \toprule
& & & & \textbf{Accrual duration} & \textbf{Trial duration} \\
\textbf{Trial Design} & \textbf{RR, Description} & \textbf{d} & \textbf{n} & \textbf{(months)} & \textbf{(months)} \\
    \midrule
         CM-017 & 1:1, Protocol & 133 & 186 & 8.5 & 21.7 \\
         Alt 1 & 3:2, Prolonged trial & 134 & (186) & (8.5) & 23.0 \\
         Alt 2 & 3:2, Accelerated accrual & 134 & (186) & 6.1 & (21.7) \\
         Alt 3 & 3:2, Increased enrollment & 134 & 196 & 8.9 & (21.7) \\
         Alt 4 & 2:1, Prolonged trial & 142 & (186) & (8.5) & 26.6 \\
         Alt 5 & 2:1, Accelerated accrual & 142 & (186) & 0 & 22.2 \\
         Alt 6 & 2:1, Increased enrollment & 142 & 210 & 9.5 & (21.7) \\
    \bottomrule
    \end{tabular}
\end{table*}
\end{center}

We consider six alternative designs (Alt 1-6 in Table \ref{tab:checkmate}). The designs differ from the Checkmate-017 protocol by randomization ratio (3:2 or 2:1), sample size, accrual duration, and/or trial duration. The latter 3 design parameters are defined according to the following ``edge cases'':
\begin{itemize}
    \item Prolonged trial: assume unequal randomization will have no impact on enrollment rate (22 patients/month). Under the alternative hypothesis, patients receiving experimental therapy will have improved survival. Thus, randomizing patients 3:2 or 2:1 may result in more time needed to observe the required number of events. \red{This edge case addresses the question, ``What is the worst delay in trial readout one can expect from employing unequal randomization (the worst case being that unequal randomization does not make the trial more attractive, hence enrollment rate stays the same)?''}
    \item Accelerated accrual: assume unequal randomization will accelerate enrollment to such a rate that the original trial timeline of $\sim$21.7 months is unaffected. \red{This edge case addresses the question, ``If the same number of patients is to be enrolled, then how much faster does the enrollment need to be for a trial under unequal randomization to avoid any delay in trial readout?''}
    \item Increased enrollment: assume unequal randomization will have no impact on enrollment rate (22 patients/month), but additional patients are enrolled at such numbers that the original trial timeline of $\sim$21.7 months is unaffected. \red{This edge case addresses the question, ``If unequal randomization does not make the trial more attractive for patient participation (and hence enrollment rate stays the same), then how many additional patients need to be enrolled for a trial under unequal randomization to avoid any delay in trial readout?''} 
\end{itemize}
Survival and dropout are assumed to be the same as in the trial protocol. For each design, 10,000 simulations are performed to calculate the number of events required to maintain 80\% power and to estimate the average trial duration.

From Table \ref{tab:checkmate}, we see that randomizing patients 3:2 (Alt 1-3) requires one additional event to maintain 80\% power when compared to the original trial protocol (CM-017). If unequal randomization does not accelerate enrollment, then trial duration can be expected to increase by 1.3 months (6.0\%) from 21.7 to 23.0 months (Alt 1). Such a delay in trial readout can be avoided if unequal randomization makes enrollment attractive, shortening accrual duration by 2.4 months (28.2\%) from 8.5 to 6.1 months---or equivalently, increasing enrollment rate by 39.6\% from 22.0 to 30.7 patients/month (Alt 2). Alternatively, trial delay can also be avoided if sample size is increased by 9 (6.8\%) from 133 to 142 patients (Alt 3). Increases in enrollment rate or sample size greater (less) than that in Alt 2 or 3 would result in a trial that is shorter (longer) than 21.7 months.

Randomizing patients 2:1 (Alt 4-6), on the other hand, requires nine additional events to maintain 80\% power. It may also increase trial duration by 4.9 months (22.6\%) from 21.7 to 26.6 months (Alt 4). This delay can be avoided if sample size is increased by 24 (12.9\%) from 186 to 210 patients (Alt 6). However, if sample size is not increased, then the trial will experience a delayed readout no matter how fast enrollment is; even when enrollment is instantaneous, the trial is expected to take 22.2 months to complete (Alt 5). 

Note that event sizes in Table \ref{tab:checkmate} were determined empirically via simulations. Had Rubinstein's equation been used instead, similar event sizes would have been obtained (134 for 3:2, 141 for 2:1). However, had Schoenfeld's equation been used, notably larger event sizes would have been suggested to maintain 80\% power (138 for 3:2, 149 for 2:1), which in turn would have exaggerated operating characteristics of the alternative trial designs (i.e., longer trial duration, larger sample size or higher enrollment rates needed to avoid delay). 

\subsection{More general cases}\label{sec:sim}
We now extend our exercise to cover a wider range of scenarios: hazard ratio (HR) 0.5, 0.6, 0.7, and 0.8; control median (CM) survival 6, 12, and 24 months; and event-patient ratio (EPR) 0.5, 0.6, 0.7, and 0.8. For each ``original'' trial design employing 1:1 randomization and defined by a unique triplet (HR, CM, EPR), we use the same combination of event size ($d$), sample size ($n$), and accrual rate as was done for the simulations described in Section \ref{sec:accuracy}. We assume that dropout follows an exponential distribution such that patients have a 1\% probability of dropping out every 12 months. Finally, we identify six alternative trial designs by following the approach in Section \ref{sec:case} (``3:2, Prolonged trial'', ``3:2, Accelerated accrual'', etc.). Results across all scenarios are summarized in Table \ref{tab:cases}. (For case-by-case results, see Tables S1--S3.)

\begin{center}
\begin{table*}[t]
    \centering
    \caption{Comparing number of events and trial duration under 3:2 and 2:1 versus 1:1 randomization in trials with 80\% power and various hazard ratios (HR), control medians (CM), and event-patient ratios (EPR). Since under the alternative hypothesis, unequal randomization (without increasing accrual rate or number of patients) leads to longer trials, we also consider the required change in accrual rate (without increasing number of patients) or number of patients (without increasing accrual rate) to avoid any trial delay.}
    \label{tab:cases}
    \begin{tabular}{p{1.5in}p{2.5in}p{2.5in}}
    \toprule
     & \multicolumn{1}{c}{\textbf{3:2 vs. 1:1}} & \multicolumn{1}{c}{\textbf{2:1 vs. 1:1}} \\ \midrule
    No. of events & Minor difference (<$10\%$) & Minor difference (<$10\%$), except when treatment effect is weaker and EPR is large (HR $\geq 0.7$, EPR $\geq 0.8$). In this case, no. of events increases >10\%. \\
    \cellcolor[gray]{0.95}Trial duration assuming same accrual rate and no. of patients & \cellcolor[gray]{0.95}Minor increase (<$10\%$), except when treatment effect is strong, survival is long, and EPR is large (HR $\leq 0.5$, CM $\geq 12$, EPR $\geq 0.8$). In this case, trial duration increases >10\%. & \cellcolor[gray]{0.95}Minor increase (<$10\%$) when EPR is small (EPR $\leq0.6$). Increase >$10\%$ otherwise.\\
    --- Accrual rate to avoid increase in trial duration & Moderate increase (<$25\%$) when survival is short or EPR is small (CM $\leq6$ or EPR $\leq0.6$). Increase >$25\%$ otherwise. & Increase >$25\%$, except when survival is short and EPR is small (CM $\leq6$, EPR $\leq0.6$). In this case, accrual rate increase is moderate (<25\%). \\
    \cellcolor[gray]{0.95}--- No. of patients to avoid increase in trial duration & \cellcolor[gray]{0.95}Minor increase (<$10\%$) & \cellcolor[gray]{0.95}Increase >$10\%$, except when treatment effect is strong and EPR is small (HR $\leq 0.6$, EPR $ \leq 0.7$). In this case, increase no. of patients is minor (<10\%).  \\
    \bottomrule
    \end{tabular}
\end{table*}
\end{center}

Compared to 1:1, 3:2 requires similar number of events to achieve 80\% power. In most cases, 3:2 leads to a minor increase in trial duration (assuming the same accrual rate and number of patients), which can be avoided by a moderate increase in accrual rate or minor increase in number of patients. A notable exception is when survival is long and EPR is large (CM $\geq12$, EPR $\geq0.7$). In this case, a >25\% increase in accrual rate is needed to avoid any trial delay. Trial delay may even be impossible to avoid in some cases; when EPR is 0.8, no amount of increase in accrual rate may offset the slower rate of accumulating events under 3:2 randomization.

2:1 also requires similar number of events as 1:1 in many cases. However, there are some cases when 2:1 requires a >10\% increase in number of events (e.g., HR=0.8).  2:1 leads to >10\% increase in trial duration when EPR is large (EPR $\geq 0.7$). Trial delay can be difficult to avoid; in many cases, a >25\% increase in accrual rate or >10\% increase in number of patients is needed. And more often than 3:2, trial delay may be unavoidable. Thus, 2:1 is more likely than 3:2 to result in a notable increase in trial duration, and a greater increase in accrual rate or number of patients is needed to avoid it.

It is worth noting that, in some cases where HR and EPR are small, we do see unequal randomization actually requiring less events than 1:1 randomization. However, the decrease in number of events is always small---2 at most. Therefore, consistent with our observation from Figure \ref{fig:optimal_rr}, 1:1 randomization may not maximize power, but it experiences little power loss compared to 3:2 and 2:1 randomization.

As an important final aside, the fact that Schoenfeld's approximation is less accurate under unequal randomization may cause us to wonder whether we should stop conducting event-driven survival trials all together; part of the lure of this longstanding practice has been the thought that power depends on the underlying HR, but not some of the other design parameters. We see from Tables S1--S3 that indeed, for trial designs under 3:2 and 2:1 randomization, the required number of events (and therefore power) depends to some extent on the control median survival and EPR. \red{However, we also see that this dependency is relatively weak, much weaker compared to the dependency on the hazard ratio} (i.e., changes to $d$ as a function of CM or EPR are much smaller compared to changes to $d$ as a function of HR).

Our recommendation is therefore two-fold. First, randomized survival trials with LRT as the primary analysis should continue to be event-driven. Second, for trials employing unequal randomization, the required number of events should be calculated \red{using a formula-based approach (e.g., Rubinstein or Schoenfeld) as a first iteration only, then subsequently fine-tuned using empirical simulations.\cite{Abel2015, rpact} Alternatively, accurate sample size formulae that rely on numerical integration (and that were originally proposed to accommodate for more complex settings such as non-exponential survival, non-proportional hazards, and treatment switching) can also be considered.\cite{Yung2020, Lakatos1988, Lu2021, Tang2021, Tang2022} Our recommendation is consistent with the current practice for trials that anticipate non-proportional hazards (NPH) at the time of trial design. Even though it is well known that under NPH, power depends on many design parameters, these trials remain event-driven. How some have instead adapted is by considering potential NPH when calculating the required number of events (see Checkmate-274\cite{Bajorin2021} and Keynote-966\cite{Kelley2023} for example).}

\section{Discussion}\label{sec:discuss}
In this paper, we revisited the question of what RR maximizes power in event-driven survival trials under PH. Our journey led us from comparing three approximations for the LRT (Section \ref{sec:logrank}), to considering other factors besides power that might influence the choice of RR (Section \ref{sec:factors}), and finally to generating insights from simulations of when 3:2 or 2:1 randomization might be of practical interest (Section \ref{sec:case}). Here is a list of summary points:
\begin{enumerate}
    \item For event-driven survival trials with HR between 0.5 and 2, Schoenfeld's approximation remains accurate under 1:1 randomization. However, Schoenfeld's approximation can be inaccurate under unequal randomization; it can underestimate power of the LRT by as much as 4\% absolute error, e.g., when HR and EPR are both small. Likewise, Freedman's approximation may over or underestimate power by 3\%. Only Rubinstein's approximation consistently estimates empirical power within 1\% (Section \ref{sec:accuracy}). 
    \item For event-driven survival trials under PH, the RR that maximizes power of the LRT is the RR that balances number of events across treatment arms at the end of the trial. \red{This was shown to be true for survival distributions under general PH (Sections \ref{sec:accuracy}-\ref{sec:generalizing_rubinstein})}.
    \item 1:1 randomization may not maximize power under PH, but compared to the RR that does, the decrease in power from employing 1:1 is often inconsequential (Figure \ref{fig:optimal_rr}). Likewise, compared to 3:2 and 2:1 randomization, 1:1 requires a few additional events at most to maintain the same level of power (Section \ref{sec:sim}). All this is to say that other reasons besides power should be considered when deciding to employ equal or unequal randomization, e.g., accrual duration, trial duration, sample size, balance in baseline characteristics, operating characteristics for estimates of secondary estimands, trial costs, individual vs. collective ethics (Section \ref{sec:factors}).
    \item In general, 3:2 and 2:1 randomization require similar number of events as 1:1. 3:2 randomization may lead to a minor increase in trial duration if accrual rate remains the same. This increase can be mitigated by a moderate increase in accrual rate or minor increase in number of patients. In contrast, 2:1 randomization may lead to a more noticeable increase in trial duration, an increase that requires greater accrual rate or number of patients to avoid. In some cases, delay in trial readout is impossible to avoid (Section \ref{sec:sim}).
    \item Although Schoenfeld's approximation is inaccurate under unequal randomization, we recommend that the practice of conducting event-driven survival trials to continue, with the caveat that the required number of events be carefully calculated (Section \ref{sec:sim}). \red{This is consistent with current practice in survival trials that anticipate NPH, even though it is known in these situations that power depends on multiple design parameters in addition to the number of events.}
\end{enumerate}

Point 1 has potential, important ramifications beyond sample size and power calculations under PH. Indeed, Schoenfeld's approximation has been widely used in other settings: trial design under non-proportional hazards\cite{Mukhopadhyay2020}, adaptive designs such as event size re-estimation\cite{Wassmer2016}, safety monitoring of overall survival\cite{Fleming2024}, assurance calculation\cite{Ren2014}, and cost-efficiency analysis\cite{Sandoval2021}, to name a few. It is therefore important in these settings to understand the potential shortcomings of Schoenfeld's approximation and to offer a more generalizable solution if necessary. 

Sample size and power calculations are more complicated under NPH compared to under PH, because they are (even) more sensitive to design assumptions such as survival distributions and EPR. A future direction for research could be coming up with guidelines or considerations for RR under specific types of NPH, e.g., delayed treatment effect, crossing hazards, and diminishing treatment effects over time.
 
Point 2 may be informative beyond the classical case that we have considered here with a two-arm randomized survival trial and a single readout. For example, we may plan differently for or have a better understanding of results from a group sequential trial, which has multiple opportunities for readout and potentially different balance of events across treatment arms over time. Likewise, we may plan differently for or have a better understanding of results from a hybrid control trial, which may employ unequal randomization for ethical or efficiency reasons, while borrowing information from historical controls if they are consistent with concurrent controls.

Finally, Point 4 should not be interpreted by the reader as encouraging or discouraging unequal randomization. \red{Our hope for this paper is rather to raise considerations that trial sponsors may have missed in the past so that more accurate assessments, efficient team discussions, and robust decision making can take place during trial design. For example, if there are ethical concerns in assigning patients to the control arm, then with the understanding that unequal randomization requires a similar number of events as 1:1 randomization, there might be greater incentive now to randomize patients 3:2 or 2:1. Alternatively, if there are cost and safety concerns for assigning more patients to the experimental arm than necessary, then with the understanding that unequal randomization prolongs the trial, there might be greater incentive for 1:1 randomization. Regardless of how the final decision is made, choosing} 
a RR in practice requires careful conversations, conversations in which statisticians are well-positioned to contribute.


\subsection*{Author contributions}

G.Y. conceived of the presented idea, developed the theory, performed the simulations, and took the lead in writing the manuscript. K.R., M.W., R.L., and Y.L. provided critical feedback and helped shape the research, analysis, and manuscript.

\subsection*{Financial disclosure}

None reported.

\subsection*{Conflict of interest}

The authors declare no potential conflict of interests.

\section*{Supporting information}

The following supporting information is available as part of the online article:

\noindent
\textbf{Figure S1.}
{Difference in power between 3:2 and 2:1 randomization.}

\noindent
\textbf{Figure S2.}
{Power vs. Randomization ratio, illustrated in more scenarios.}

\noindent
\textbf{Table S1.}
{`Unequal randomization, Prolonged trial' vs. `Equal randomization', case-by-case results.}

\noindent
\textbf{Table S2.}
{`Unequal randomization, Accelerated accrual' vs. `Equal randomization', case-by-case results.}

\noindent
\textbf{Table S3.}
{`Unequal randomization, Increased enrollment' vs. `Equal randomization', case-by-case results.}

\appendix

\section{Ratio of expected number of events} \label{app1}
In addition to the notations in Section \ref{sec:logrank}, let $\wedge$ denote the minimum, $a(\cdot)=A'(\cdot)$ the probability density function (PDF) for enrollment, $f_e(y)=-\frac{d}{dt}S_e(t)$ the PDF for event, and $\overline{L}(\cdot)=1-L(\cdot)$ the complement of the lost-to-followup distribution. We assume uniform enrollment, exponential dropout with hazard $\eta$, and exponential survival with hazard $\lambda_e$. The expected number of patients on the experimental arm that have experienced an event by time $t$ after first patient in is given by
\begin{align*}
E(D_e(t)) 
    &= n \pi \int_0^{r \wedge t} a(x) \int_0^{t-x} f_e(y) \overline{L}(y) dy dx \\
    &= n \pi \int_0^{r \wedge t} \frac{1}{r} \int_0^{t-x} \lambda_e e^{-\lambda_e y} e^{-\eta y} dy dx \\
    &= \frac{n \pi}{r} \left \{ \frac{\lambda_e}{\lambda_e+\eta} (r \wedge t) - \frac{\lambda_e}{(\lambda_e+\eta)^2} e^{(\lambda_e + \eta)((r-t) \wedge 0)} + \frac{\lambda_e}{(\lambda_e+\eta)^2} e^{-(\lambda_e + \eta)t} \right \}.
\end{align*} 
Likewise, the expected number of patients on the control arm that have experienced an event by time $t$ is given by
\begin{equation*}
    E(D_c(t)) = \frac{n(1-\pi)}{r} \left \{ \frac{\lambda_c}{\lambda_c+\eta} (r \wedge t) - \frac{\lambda_c}{(\lambda_c+\eta)^2} e^{(\lambda_c + \eta)((r-t) \wedge 0)} + \frac{\lambda_c}{(\lambda_c+\eta)^2} e^{-(\lambda_c + \eta)t} \right \}.
\end{equation*}
It is easy to show that
\begin{equation*}
    \lim_{t \rightarrow \infty} \frac{E(D_e(t))}{E(D_c(t))} = \frac{\pi}{1-\pi} \cdot \frac{\lambda_e / (\lambda_e + \eta)}{\lambda_c / (\lambda_c + \eta)}.
\end{equation*}
Applying L'Hopital's Rule twice, one can also show that
\begin{equation*}
    \lim_{t \rightarrow 0^+} \frac{E(D_e(t))}{E(D_c(t))} = \frac{\pi}{1-\pi} \cdot \frac{\lambda_e}{\lambda_c}.
\end{equation*}

\bibliography{main}

\begin{thebibliography}{10}
\providecommand \doibase [0]{http://dx.doi.org/}%

\bibitem{Schoenfeld1981}
Schoenfeld D. The asymptotic properties of nonparametric tests for comparing
  survival distributions. {\it Biometrika} 1981\string; 68\string: 316-319.

\bibitem{Pocock1979}
Pocock SJ. Allocation of patients to treatment in clinical trials. {\it
  Biometrics} 1979\string; 35\string: 183-197.

\bibitem{Avins1998}
Avins AL. Can unequal be more fair? Ethics, subject allocation, and randomised
  clinical trials. {\it Journal of Medical Ethics} 1998\string; 24\string:
  401-408.

\bibitem{Dumville2006}
Dumville JC, Hahn S, Miles JN, Torgerson DJ. The use of unequal randomisation
  ratios in clinical trials: A review. {\it Contemporary Clinical Trials}
  2006\string; 27\string: 1-12.

\bibitem{Hey2014}
Hey SP, Kimmelman J. The questionable use of unequal allocation in confirmatory
  trials. {\it Neurology} 2014\string; 82\string: 77-79.

\bibitem{Peckham2015}
Peckham E, Brabyn S, Cook L, Devlin T, Dumville J, Torgerson DJ. The use of
  unequal randomisation in clinical trials - An update. {\it Contemporary
  Clinical Trials} 2015\string; 45\string: 113-122.

\bibitem{Rubinstein1981}
Rubinstein LV, Gail MH, Santner TJ. Planning the duration of a comparative
  clinical trial with loss to follow-up and a period of continued observation.
  {\it J Chron Dis} 1981\string; 34\string: 469-479.

\bibitem{Sposto1987}
Sposto R, Krailo MD. Use of unequal allocation in survival trials. {\it
  Statistics in Medicine} 1987\string; 6\string: 119-125.

\bibitem{Freedman1982}
Freedman LS. Tables of the number of patients required in clinical trials using
  the logrank test. {\it Statistics in Medicine} 1982\string; 1\string:
  121-129.

\bibitem{Hsieh1992}
Hsieh FY. Comparing sample size formulae for trials with unbalanced allocation
  using the logrank test. {\it Statistics in Medicine} 1992\string; 11\string:
  1091-1098.

\bibitem{Abel2015}
Abel UR, Jensen K, Karapanagiotou-Schenkel I, Kieser M. Some Issues of Sample
  Size Calculation for Time-to-Event Endpoints Using the Freedman and
  Schoenfeld Formulas. {\it Journal of Biopharmaceutical Statistics}
  2015\string; 25\string: 1285-1311.

\bibitem{Yung2020}
Yung G, Liu Y. Sample size and power for the weighted log-rank test and
  Kaplan-Meier based tests with allowance for nonproportional hazards. {\it
  Biometrics} 2020\string; 76\string: 939-950.

\bibitem{Sandoval2021}
Sandoval GJ, Bebu I, Lachin JM. Cost-efficient clinical studies with continuous
  time survival outcomes. {\it Statistics in Medicine} 2021\string; 40\string:
  3682-3694.

\bibitem{Peto1972}
Peto R, Peto J. Asymptotically efficient rank invariant test procedures. {\it
  Journal of the Royal Statistical Society Series A} 1972\string; 135\string:
  185-207.

\bibitem{Barthel2006}
Barthel FM, Babiker A, Royston P, Parmar MK. Evaluation of sample size and
  power for multi-arm survival trails allowing for non-uniform accrual,
  non-proportional hazards, loss to follow-up and cross-over. {\it Statistics
  in Medicine} 2006\string; 25\string: 2521-2542.

\bibitem{Wu2015}
Wu J. Power and sample size for randomized phase III survival trials under the
  weibull model. {\it Journal of Biopharmaceutical Statistics} 2015\string;
  25\string: 16-28.

\bibitem{Fay2018}
Fay MP, Malinovsky Y. Confidence intervals of the Mann-Whitney parameter that
  are compatible with the Wilcoxon-Mann-Whitney test. {\it Statistics in
  Medicine} 2018\string; 37\string: 3991-4006.

\bibitem{Eek2016}
Eek D, Krohe M, Mazar I, et al. Patient-reported preferences for oral versus
  intravenous administration for the treatment of cancer: A review of the
  literature. {\it Patient Preference and Adherence} 2016\string; 10\string:
  1609-1621.

\bibitem{Nipp2019}
Nipp RD, Hong K, Paskett ED. Overcoming barriers to clinical trial enrollment.
  {\it ASCO Educational Book} 2019\string; Jan\string: 105-114.

\bibitem{Wong2020}
Wong AR, Sun V, George K, et al. Barriers to participation in therapeutic
  clinical trials as perceived by community oncologists. {\it JCO Oncology
  Practice} 2020\string; 16\string: e849-e858.

\bibitem{Roberts1999}
Roberts C, Torgerson DJ. Baseline imbalance in randomised controlled trials.
  {\it British Medical Journal} 1999\string; 319\string: 185.

\bibitem{ICH2019}
ICH . Addendum on estimands and sensitivity analysis in clinical trials to the
  guideline on statistical principles for clinical trials.  2019\string;
  E9(R1).

\bibitem{Clark2002}
Clark TG, Altman DG, Stavola BLD. Quantification of the completeness of
  follow-up. {\it The Lancet} 2002\string; 359\string: 1309-1310.

\bibitem{Pocock2002}
Pocock SJ. Survival plots of time-to-event out practice and pitfalls. {\it The
  Lancet} 2002\string; 359\string: 1686-1689.

\bibitem{Betensky2015}
Betensky RA. Measures of follow-up in time-to-event studies: Why provide them
  and what should they be?. {\it Clinical Trials} 2015\string; 12\string:
  403-408.

\bibitem{Rufibach2023}
Rufibach K, Grinsted L, Li J, Weber HJ, Zheng C, Zhou J. Quantification of
  follow-up time in oncology clinical trials with a time-to-event endpoint:
  Asking the right questions. {\it Pharmaceutical Statistics} 2023\string;
  22\string: 671-691.

\bibitem{Fleming2024}
Fleming TR, Hampson LV, Bharani BD, et al. Monitoring overall survival in
  pivotal trials in indolent cancers. {\it Statistics in Biopharmaceutical
  Research} 2024\string: 1-20.

\bibitem{Torgerson2000}
Torgerson DJ, Campbell MK. Use of unequal randomisation to aid the economic
  efficiency of clinical trials. {\it British Medical Journal} 2000\string;
  321\string: 759.

\bibitem{Chandereng2020}
Chandereng T, Wei X, Chappell R. Imbalanced randomization in clinical trials.
  {\it Statistics in Medicine} 2020\string; 39\string: 2185-2196.

\bibitem{Senn2002}
Senn S. Ethical considerations concerning treatment allocation in drug
  development trials. {\it Statistical Methods in Medical Research}
  2002\string; 11\string: 403-411.

\bibitem{Brahmer2015}
Brahmer J, Reckamp KL, Baas P, et al. Nivolumab versus Docetaxel in Advanced
  Squamous-Cell Non–Small-Cell Lung Cancer. {\it New England Journal of
  Medicine} 2015\string; 373\string: 123-135.

\bibitem{rpact}
Wassmer G, Pahlke F. rpact: Confirmatory Adaptive Clinical Trial Design and
  Analysis. R package version 4.0.0.;  2024.
\newblock https://CRAN.R-project.org/package=rpact.

\bibitem{Lakatos1988}
Lakatos E. Sample Sizes Based on the Log-Rank Statistic in Complex Clinical
  Trials. {\it Biometrics} 1988\string; 44\string: 229-241.

\bibitem{Lu2021}
Lu K. Sample size calculation for logrank test and prediction of number of
  events over time. {\it Pharmaceutical Statistics} 2021\string; 20\string:
  229-244.

\bibitem{Tang2021}
Tang Y. A unified approach to power and sample size determination for log-rank
  tests under proportional and nonproportional hazards. {\it Statistical
  Methods in Medical Research} 2021\string; 30\string: 1211-1234.

\bibitem{Tang2022}
Tang Y. Complex survival trial design by the product integration method. {\it
  Statistics in Medicine} 2022\string; 41\string: 798-814.

\bibitem{Bajorin2021}
Bajorin DF, Witjes JA, Gschwend JE, et al. Adjuvant Nivolumab versus Placebo in
  Muscle-Invasive Urothelial Carcinoma. {\it New England Journal of Medicine}
  2021\string; 384\string: 2102-2114.

\bibitem{Kelley2023}
Kelley RK, Ueno M, Yoo C, et al. Pembrolizumab in combination with gemcitabine
  and cisplatin compared with gemcitabine and cisplatin alone for patients with
  advanced biliary tract cancer (KEYNOTE-966): a randomised, double-blind,
  placebo-controlled, phase 3 trial. {\it The Lancet} 2023\string; 401\string:
  1853-1865.

\bibitem{Mukhopadhyay2020}
Mukhopadhyay P, Huang W, Metcalfe P, Öhrn F, Jenner M, Stone A. Statistical
  and practical considerations in designing of immuno-oncology trials. {\it
  Journal of Biopharmaceutical Statistics} 2020\string; 30\string: 1130-1146.

\bibitem{Wassmer2016}
Wassmer G, Brannath W. {\it Group sequential and confirmatory adaptive designs
  in clinical trials}.
\newblock Springer Cham .
\newblock 2016.

\bibitem{Ren2014}
Ren S, Oakley JE. Assurance calculations for planning clinical trials with
  time-to-event outcomes. {\it Statistics in Medicine} 2014\string; 33\string:
  31-45.

\end{thebibliography}

\clearpage



\end{document}


\maketitle

\begin{figure}
    \centering
    \includegraphics[width=0.8\textwidth]{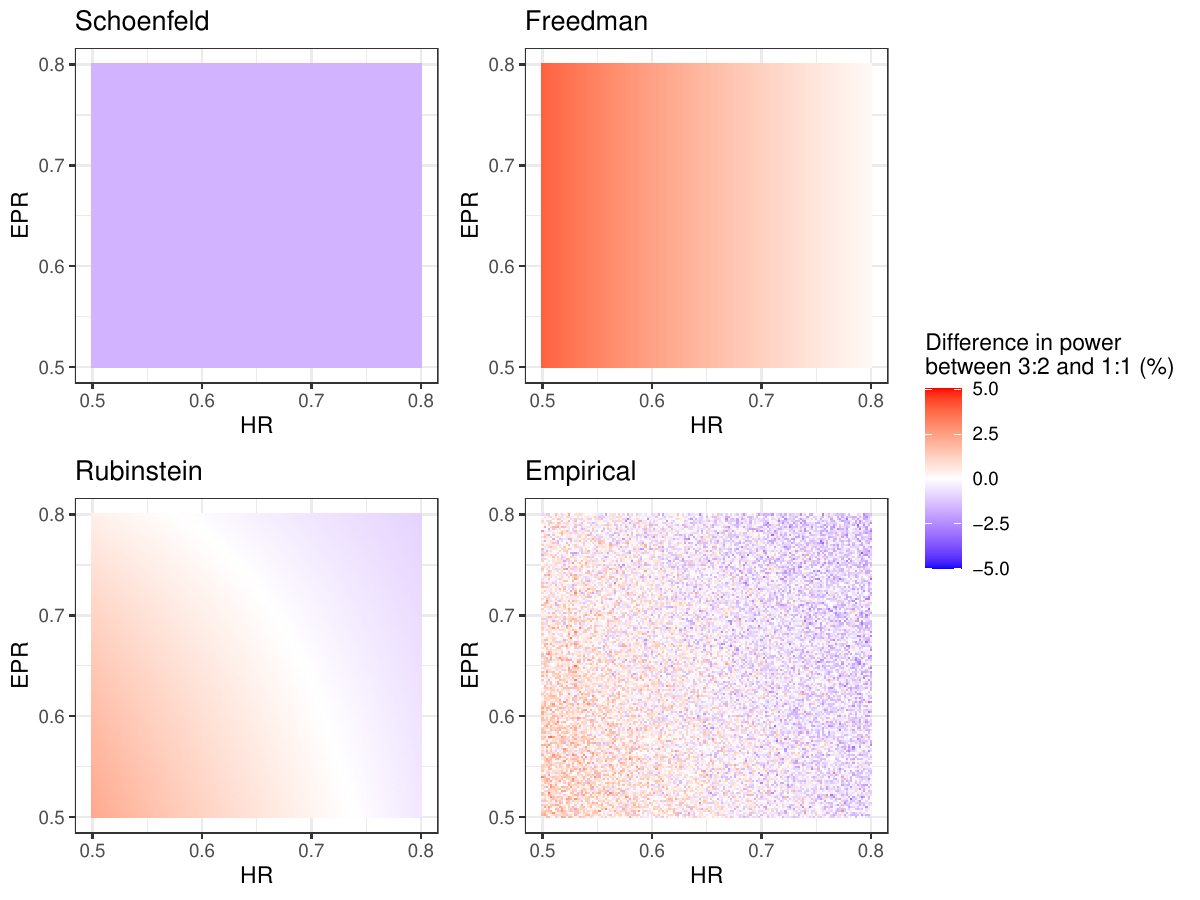}
    \caption{Difference in power between 3:2 and 1:1 randomization, according to various approximations or empirical simulations and across scenarios with different hazard ratios (HR) and event-patient ratios (EPR).}
    \label{fig:accuracy2}
\end{figure}

\begin{figure}
    \centering
    \includegraphics[width=0.8\textwidth]{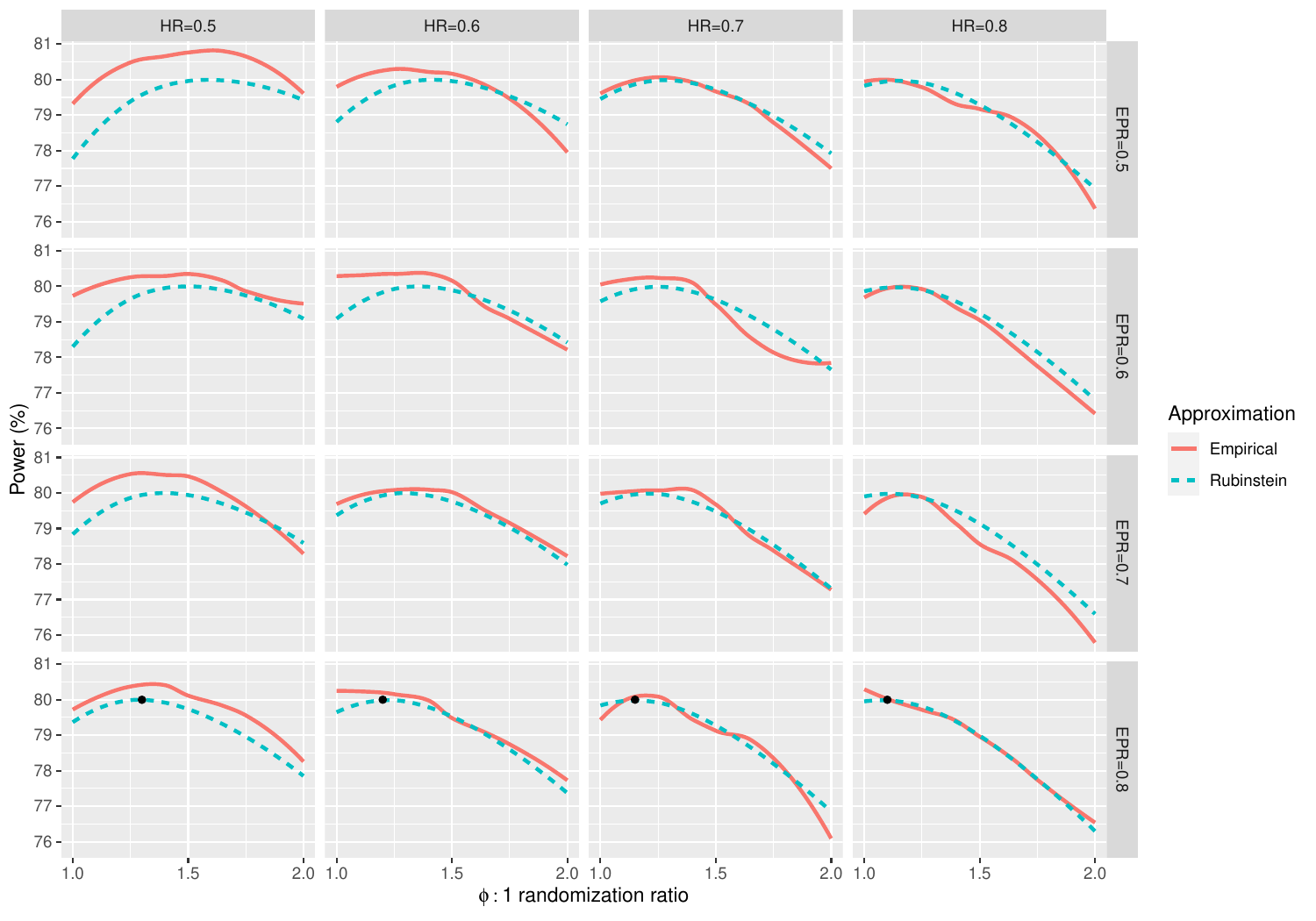}
    \caption{Power for various randomization ratios, according to 10,000 empirical simulations and Rubinstein's approximation. 16 scenarios with varying hazard ratios (HR) and event-patient ratios (EPR) are considered. Black dots indicate the point at which power is maximized under Rubinstein's approximation. They also indicate the randomization ratio under which events are expected to be balanced across treatment arms at the end of the trial. LOESS smoothing was applied to the empirical curves to more clearly display trends amidst Monte Carlo error. The Monte Carlo standard error is estimated to be $\sim$0.4\% given 10,000 simulations and true power $\sim$80\%.}
    \label{fig:accuracy2}
\end{figure}

\begin{table}
\caption{`Unequal randomization, Prolonged trial' vs. `Equal randomization'. Given a hazard ratio and control median survival (row), a default design is proposed (columns 4-6) that: employs 1:1 randomization; targets $d$ events following Schoenfeld's equation to achieve 80\% power with one-sided $\alpha=0.025$; fixes event-patient ratio ($d/n$) to approximately 0.5, 0.6, 0.7, or 0.8; and enrolls 20-50 patients per month over $r$ months. The expected total trial duration is denoted by $t_d$. For each default design, two alternative designs are proposed, one employing 3:2 randomization (columns 7-8), the second employing 2:1 randomization (columns 9-10). Both alternative designs maintain 80\% power by recalculating the target number of events $d$. Given that accrual rate and sample size are fixed, expected trial duration $t_d$ may be prolonged. $\Delta d$ and $\Delta t_d$ denote the change in number of events and expected trial duration compared to the default design under 1:1 randomization.}
\centering
\footnotesize
\begin{tabular}{rrrrrrrrrrrr}
  \hline
Hazard & Control & \multicolumn{4}{c}{1:1} & & \multicolumn{2}{c}{3:2} & & \multicolumn{2}{c}{2:1} \\
\cline{3-6} \cline{8-9} \cline{11-12}
Ratio & Median & $d$ & $n$ & $r$ & $t_d$ & & $\Delta d$ & $\Delta t_d$ & & $\Delta d$ & $\Delta t_d$ \\ \hline \hline
0.5 & 6 & 66 & 132 & 6.6 & 11.8 &  & -2 (-3\%) & 0.1 (+1\%) &  & -2 (-3\%) & 0.5 (+4\%) \\ 
   &  & 66 & 110 & 5.5 & 14.1 &  & -2 (-3\%) & 0.2 (+1\%) &  & -1 (-2\%) & 0.9 (+7\%) \\ 
   &  & 66 & 96 & 4.8 & 17.0 &  & -2 (-3\%) & 0.4 (+3\%) &  & 0 (+0\%) & 2.1 (+13\%) \\ 
   &  & 66 & 84 & 4.2 & 22.0 &  & -1 (-2\%) & 1.3 (+6\%) &  & 3 (+5\%) & 6.3 (+29\%) \\ 
   & 12 & 66 & 132 & 6.6 & 20.2 &  & -2 (-3\%) & 0.3 (+1\%) &  & -2 (-3\%) & 1.1 (+5\%) \\ 
   &  & 66 & 110 & 5.5 & 25.4 &  & -2 (-3\%) & 0.4 (+2\%) &  & -1 (-2\%) & 1.9 (+7\%) \\ 
   &  & 66 & 96 & 4.8 & 31.8 &  & -2 (-3\%) & 0.9 (+3\%) &  & 1 (+2\%) & 5.5 (+17\%) \\ 
   &  & 66 & 84 & 4.2 & 42.4 &  & 0 (+0\%) & 4.6 (+11\%) &  & 3 (+5\%) & 13.3 (+31\%) \\ 
   & 24 & 66 & 132 & 6.6 & 37.3 &  & -2 (-3\%) & 0.6 (+2\%) &  & -2 (-3\%) & 2.1 (+6\%) \\ 
   &  & 66 & 110 & 5.5 & 48.7 &  & -2 (-3\%) & 0.9 (+2\%) &  & -1 (-2\%) & 3.9 (+8\%) \\ 
   &  & 66 & 96 & 4.8 & 62.4 &  & -2 (-3\%) & 1.9 (+3\%) &  & 1 (+2\%) & 11.6 (+19\%) \\ 
   &  & 66 & 84 & 4.2 & 85.5 &  & 0 (+0\%) & 10 (+12\%) &  & 4 (+6\%) & 35.8 (+42\%) \\ 
  0.6 & 6 & 121 & 242 & 8.1 & 12.0 &  & -2 (-2\%) & 0.2 (+1\%) &  & 2 (+2\%) & 0.9 (+7\%) \\ 
   &  & 121 & 202 & 6.7 & 13.8 &  & -2 (-2\%) & 0.2 (+2\%) &  & 3 (+2\%) & 1.3 (+10\%) \\ 
   &  & 121 & 174 & 5.8 & 16.5 &  & -1 (-1\%) & 0.5 (+3\%) &  & 5 (+4\%) & 2.5 (+15\%) \\ 
   &  & 121 & 152 & 5.1 & 21.0 &  & 0 (+0\%) & 1 (+5\%) &  & 8 (+7\%) & 5.7 (+27\%) \\ 
   & 12 & 121 & 242 & 8.1 & 19.6 &  & -2 (-2\%) & 0.4 (+2\%) &  & 1 (+1\%) & 1.5 (+8\%) \\ 
   &  & 121 & 202 & 6.7 & 24.1 &  & -2 (-2\%) & 0.5 (+2\%) &  & 3 (+2\%) & 2.7 (+11\%) \\ 
   &  & 121 & 174 & 5.8 & 30.2 &  & -1 (-1\%) & 1 (+3\%) &  & 5 (+4\%) & 5.2 (+17\%) \\ 
   &  & 121 & 152 & 5.1 & 39.9 &  & 0 (+0\%) & 2 (+5\%) &  & 9 (+7\%) & 13.4 (+33\%) \\ 
   & 24 & 121 & 242 & 8.1 & 35.3 &  & -2 (-2\%) & 0.7 (+2\%) &  & 1 (+1\%) & 3.1 (+9\%) \\ 
   &  & 121 & 202 & 6.7 & 45.2 &  & -2 (-2\%) & 1 (+2\%) &  & 3 (+2\%) & 5.5 (+12\%) \\ 
   &  & 121 & 174 & 5.8 & 58.4 &  & -1 (-1\%) & 2 (+3\%) &  & 5 (+4\%) & 10.8 (+19\%) \\ 
   &  & 121 & 152 & 5.1 & 79.6 &  & 0 (+0\%) & 4.3 (+5\%) &  & 9 (+7\%) & 29.8 (+37\%) \\ 
  0.7 & 6 & 247 & 494 & 12.4 & 14.0 &  & 0 (+0\%) & 0.2 (+2\%) &  & 12 (+5\%) & 1 (+7\%) \\ 
   &  & 247 & 412 & 10.3 & 15.1 &  & 1 (+0\%) & 0.4 (+3\%) &  & 14 (+6\%) & 1.5 (+10\%) \\ 
   &  & 247 & 354 & 8.9 & 17.3 &  & 2 (+1\%) & 0.7 (+4\%) &  & 17 (+7\%) & 2.7 (+16\%) \\ 
   &  & 247 & 310 & 7.8 & 21.0 &  & 4 (+2\%) & 1.4 (+7\%) &  & 23 (+9\%) & 6.7 (+32\%) \\ 
   & 12 & 247 & 494 & 12.4 & 20.9 &  & 0 (+0\%) & 0.5 (+2\%) &  & 11 (+4\%) & 1.9 (+9\%) \\ 
   &  & 247 & 412 & 10.3 & 24.5 &  & 1 (+0\%) & 0.8 (+3\%) &  & 14 (+6\%) & 3.1 (+13\%) \\ 
   &  & 247 & 354 & 8.9 & 29.8 &  & 2 (+1\%) & 1.4 (+5\%) &  & 17 (+7\%) & 5.7 (+19\%) \\ 
   &  & 247 & 310 & 7.8 & 38.2 &  & 4 (+2\%) & 2.9 (+7\%) &  & 23 (+9\%) & 14.2 (+37\%) \\ 
   & 24 & 247 & 494 & 12.4 & 35.3 &  & 0 (+0\%) & 1 (+3\%) &  & 11 (+4\%) & 3.8 (+11\%) \\ 
   &  & 247 & 412 & 10.3 & 44.0 &  & 1 (+0\%) & 1.7 (+4\%) &  & 13 (+5\%) & 6.1 (+14\%) \\ 
   &  & 247 & 354 & 8.9 & 55.9 &  & 2 (+1\%) & 2.8 (+5\%) &  & 17 (+7\%) & 11.9 (+21\%) \\ 
   &  & 247 & 310 & 7.8 & 74.3 &  & 4 (+2\%) & 6.1 (+8\%) &  & 24 (+10\%) & 33.3 (+45\%) \\ 
  0.8 & 6 & 631 & 1262 & 25.2 & 21.3 &  & 13 (+2\%) & 0.4 (+2\%) &  & 55 (+9\%) & 1.5 (+7\%) \\ 
   &  & 631 & 1052 & 21.0 & 21.3 &  & 13 (+2\%) & 0.5 (+2\%) &  & 56 (+9\%) & 1.7 (+8\%) \\ 
   &  & 631 & 902 & 18.0 & 22.1 &  & 14 (+2\%) & 0.8 (+4\%) &  & 58 (+9\%) & 2.9 (+13\%) \\ 
   &  & 631 & 790 & 15.8 & 24.7 &  & 17 (+3\%) & 1.5 (+6\%) &  & 66 (+10\%) & 6.4 (+26\%) \\ 
   & 12 & 631 & 1262 & 25.2 & 27.5 &  & 10 (+2\%) & 0.6 (+2\%) &  & 49 (+8\%) & 2.1 (+8\%) \\ 
   &  & 631 & 1052 & 21.0 & 29.4 &  & 11 (+2\%) & 0.9 (+3\%) &  & 52 (+8\%) & 3.4 (+11\%) \\ 
   &  & 631 & 902 & 18.0 & 33.4 &  & 13 (+2\%) & 1.6 (+5\%) &  & 57 (+9\%) & 5.9 (+18\%) \\ 
   &  & 631 & 790 & 15.8 & 40.4 &  & 16 (+3\%) & 3 (+7\%) &  & 66 (+10\%) & 13.5 (+33\%) \\ 
   & 24 & 631 & 1262 & 25.2 & 40.5 &  & 8 (+1\%) & 1.1 (+3\%) &  & 47 (+7\%) & 4.2 (+10\%) \\ 
   &  & 631 & 1052 & 21.0 & 47.2 &  & 10 (+2\%) & 1.8 (+4\%) &  & 51 (+8\%) & 6.8 (+15\%) \\ 
   &  & 631 & 902 & 18.0 & 57.5 &  & 12 (+2\%) & 3.1 (+5\%) &  & 57 (+9\%) & 12.4 (+22\%) \\ 
   &  & 631 & 790 & 15.8 & 73.7 &  & 16 (+3\%) & 6.5 (+9\%) &  & 67 (+11\%) & 30.7 (+42\%) \\
\hline
\end{tabular}
\end{table}

\begin{table}
\caption{`Unequal randomization, Accelerated accrual' vs. `Equal randomization'. Given a hazard ratio and control median survival (row), a default design is proposed (columns 4-6) that: employs 1:1 randomization; targets $d$ events following Schoenfeld's equation to achieve 80\% power with one-sided $\alpha=0.025$; fixes event-patient ratio ($d/n$) to approximately 0.5, 0.6, 0.7, or 0.8; and enrolls 20-50 patients per month over $r$ months. The expected total trial duration is denoted by $t_d$. For each default design, two alternative designs are proposed, one employing 3:2 randomization (columns 7-8), the second employing 2:1 randomization (columns 9-10). Both alternative designs maintain 80\% power by recalculating the target number of events $d$. Expected trial duration $t_d$ is also maintained, if possible, by accelerating accrual. $\Delta d$ and $\Delta r$ denote the change in number of events and accrual period compared to the default design under 1:1 randomization.}
\centering
\footnotesize
\begin{tabular}{rrrrrrrrrrrr}
  \hline
Hazard & Control & \multicolumn{4}{c}{1:1} & & \multicolumn{2}{c}{3:2} & & \multicolumn{2}{c}{2:1} \\
\cline{3-6} \cline{8-9} \cline{11-12}
Ratio & Median & $d$ & $n$ & $r$ & $t_d$ & & $\Delta d$ & $\Delta r$ & & $\Delta d$ & $\Delta r$ \\ \hline \hline
0.5 & 6 & 66 & 132 & 6.6 & 11.8 &  & -2 (-3\%) & -0.2 (-4\%) &  & -2 (-3\%) & -0.9 (-14\%) \\ 
   &  & 66 & 110 & 5.5 & 14.1 &  & -2 (-3\%) & -0.5 (-9\%) &  & -1 (-2\%) & -1.7 (-32\%) \\ 
   &  & 66 & 96 & 4.8 & 17.0 &  & -2 (-3\%) & -0.9 (-20\%) &  & 0 (+0\%) & -4.2 (-89\%) \\ 
   &  & 66 & 84 & 4.2 & 22.0 &  & -1 (-2\%) & -2.7 (-63\%) &  & -2 (-3\%) & NA \\ 
   & 12 & 66 & 132 & 6.6 & 20.2 &  & -2 (-3\%) & -0.5 (-8\%) &  & -2 (-3\%) & -2 (-31\%) \\ 
   &  & 66 & 110 & 5.5 & 25.4 &  & -2 (-3\%) & -0.9 (-16\%) &  & -1 (-2\%) & -3.7 (-68\%) \\ 
   &  & 66 & 96 & 4.8 & 31.8 &  & -2 (-3\%) & -1.8 (-39\%) &  & -2 (-3\%) & NA \\ 
   &  & 66 & 84 & 4.2 & 42.4 &  & -2 (-3\%) & NA &  & -3 (-5\%) & NA \\ 
   & 24 & 66 & 132 & 6.6 & 37.3 &  & -2 (-3\%) & -1.1 (-17\%) &  & -2 (-3\%) & -4.2 (-64\%) \\ 
   &  & 66 & 110 & 5.5 & 48.7 &  & -2 (-3\%) & -1.8 (-33\%) &  & -2 (-3\%) & NA \\ 
   &  & 66 & 96 & 4.8 & 62.4 &  & -2 (-3\%) & -4 (-82\%) &  & -4 (-6\%) & NA \\ 
   &  & 66 & 84 & 4.2 & 85.5 &  & -2 (-3\%) & NA &  & -4 (-6\%) & NA \\ 
  0.6 & 6 & 121 & 242 & 8.1 & 12.0 &  & -2 (-2\%) & -0.4 (-5\%) &  & 2 (+2\%) & -1.6 (-19\%) \\ 
   &  & 121 & 202 & 6.7 & 13.8 &  & -2 (-2\%) & -0.5 (-7\%) &  & 3 (+2\%) & -2.5 (-37\%) \\ 
   &  & 121 & 174 & 5.8 & 16.5 &  & -1 (-1\%) & -0.9 (-16\%) &  & 5 (+4\%) & -4.9 (-84\%) \\ 
   &  & 121 & 152 & 5.1 & 21.0 &  & 0 (+0\%) & -1.9 (-37\%) &  & 2 (+2\%) & NA \\ 
   & 12 & 121 & 242 & 8.1 & 19.6 &  & -2 (-2\%) & -0.7 (-8\%) &  & 1 (+1\%) & -3 (-37\%) \\ 
   &  & 121 & 202 & 6.7 & 24.1 &  & -2 (-2\%) & -1 (-15\%) &  & 3 (+2\%) & -5.2 (-77\%) \\ 
   &  & 121 & 174 & 5.8 & 30.2 &  & -1 (-1\%) & -1.9 (-33\%) &  & 0 (+0\%) & NA \\ 
   &  & 121 & 152 & 5.1 & 39.9 &  & 0 (+0\%) & -4 (-79\%) &  & -1 (-1\%) & NA \\ 
   & 24 & 121 & 242 & 8.1 & 35.3 &  & -2 (-2\%) & -1.5 (-18\%) &  & 1 (+1\%) & -6.2 (-76\%) \\ 
   &  & 121 & 202 & 6.7 & 45.2 &  & -2 (-2\%) & -2.1 (-31\%) &  & -1 (-1\%) & NA \\ 
   &  & 121 & 174 & 5.8 & 58.4 &  & -1 (-1\%) & -4 (-69\%) &  & -3 (-2\%) & NA \\ 
   &  & 121 & 152 & 5.1 & 79.6 &  & -1 (-1\%) & NA &  & -3 (-2\%) & NA \\ 
  0.7 & 6 & 247 & 494 & 12.4 & 14.0 &  & 0 (+0\%) & -0.5 (-4\%) &  & 12 (+5\%) & -1.7 (-14\%) \\ 
   &  & 247 & 412 & 10.3 & 15.1 &  & 1 (+0\%) & -0.7 (-7\%) &  & 14 (+6\%) & -2.7 (-26\%) \\ 
   &  & 247 & 354 & 8.9 & 17.3 &  & 2 (+1\%) & -1.2 (-14\%) &  & 17 (+7\%) & -5.1 (-58\%) \\ 
   &  & 247 & 310 & 7.8 & 21.0 &  & 4 (+2\%) & -2.5 (-32\%) &  & 13 (+5\%) & NA \\ 
   & 12 & 247 & 494 & 12.4 & 20.9 &  & 0 (+0\%) & -1 (-8\%) &  & 11 (+4\%) & -3.5 (-28\%) \\ 
   &  & 247 & 412 & 10.3 & 24.5 &  & 1 (+0\%) & -1.6 (-16\%) &  & 14 (+6\%) & -6 (-58\%) \\ 
   &  & 247 & 354 & 8.9 & 29.8 &  & 2 (+1\%) & -2.6 (-29\%) &  & 12 (+5\%) & NA \\ 
   &  & 247 & 310 & 7.8 & 38.2 &  & 4 (+2\%) & -5.5 (-71\%) &  & 3 (+1\%) & NA \\ 
   & 24 & 247 & 494 & 12.4 & 35.3 &  & 0 (+0\%) & -2 (-16\%) &  & 11 (+4\%) & -7.5 (-61\%) \\ 
   &  & 247 & 412 & 10.3 & 44.0 &  & 1 (+0\%) & -3.4 (-33\%) &  & 10 (+4\%) & NA \\ 
   &  & 247 & 354 & 8.9 & 55.9 &  & 2 (+1\%) & -5.6 (-63\%) &  & 2 (+1\%) & NA \\ 
   &  & 247 & 310 & 7.8 & 74.3 &  & 1 (+0\%) & NA &  & -2 (-1\%) & NA \\ 
  0.8 & 6 & 631 & 1262 & 25.2 & 21.3 &  & 13 (+2\%) & -0.8 (-3\%) &  & 55 (+9\%) & -2.5 (-10\%) \\ 
   &  & 631 & 1052 & 21.0 & 21.3 &  & 13 (+2\%) & -0.8 (-4\%) &  & 56 (+9\%) & -2.7 (-13\%) \\ 
   &  & 631 & 902 & 18.0 & 22.1 &  & 14 (+2\%) & -1.3 (-7\%) &  & 58 (+9\%) & -4.7 (-26\%) \\ 
   &  & 631 & 790 & 15.8 & 24.7 &  & 17 (+3\%) & -2.5 (-16\%) &  & 66 (+10\%) & -11 (-70\%) \\ 
   & 12 & 631 & 1262 & 25.2 & 27.5 &  & 10 (+2\%) & -1 (-4\%) &  & 49 (+8\%) & -3.6 (-14\%) \\ 
   &  & 631 & 1052 & 21.0 & 29.4 &  & 11 (+2\%) & -1.7 (-8\%) &  & 52 (+8\%) & -5.9 (-28\%) \\ 
   &  & 631 & 902 & 18.0 & 33.4 &  & 13 (+2\%) & -2.8 (-16\%) &  & 57 (+9\%) & -10.7 (-59\%) \\ 
   &  & 631 & 790 & 15.8 & 40.4 &  & 16 (+3\%) & -5.5 (-35\%) &  & 43 (+7\%) & NA \\ 
   & 24 & 631 & 1262 & 25.2 & 40.5 &  & 8 (+1\%) & -2.1 (-8\%) &  & 47 (+7\%) & -7.8 (-31\%) \\ 
   &  & 631 & 1052 & 21.0 & 47.2 &  & 10 (+2\%) & -3.4 (-16\%) &  & 51 (+8\%) & -13 (-62\%) \\ 
   &  & 631 & 902 & 18.0 & 57.5 &  & 12 (+2\%) & -5.8 (-32\%) &  & 41 (+6\%) & NA \\ 
   &  & 631 & 790 & 15.8 & 73.7 &  & 16 (+3\%) & -12.6 (-80\%) &  & 16 (+3\%) & NA \\ 
\hline
\end{tabular}
\end{table}

\begin{table}
\caption{`Unequal randomization, Increased enrollment' vs. `Equal randomization'. Given a hazard ratio and control median survival (row), a default design is proposed (columns 4-6) that: employs 1:1 randomization; targets $d$ events following Schoenfeld's equation to achieve 80\% power with one-sided $\alpha=0.025$; fixes event-patient ratio ($d/n$) to approximately 0.5, 0.6, 0.7, or 0.8; and enrolls 20-50 patients per month over $r$ months. The expected total trial duration is denoted by $t_d$. For each default design, two alternative designs are proposed, one employing 3:2 randomization (columns 7-8), the second employing 2:1 randomization (columns 9-10). Both alternative designs maintain 80\% power by recalculating the target number of events $d$. Expected trial duration $t_d$ is also maintained, if possible, by enrolling more patients (but at the same enrollment rate). $\Delta d$ and $\Delta n$ denote the change in number of events and number of patients compared to the default design under 1:1 randomization.}
\centering
\footnotesize
\begin{tabular}{rrrrrrrrrrrr}
  \hline
Hazard & Control & \multicolumn{4}{c}{1:1} & & \multicolumn{2}{c}{3:2} & & \multicolumn{2}{c}{2:1} \\
\cline{3-6} \cline{8-9} \cline{11-12}
Ratio & Median & $d$ & $n$ & $\tau_a$ & $\tau$ & & $\Delta d$ & $\Delta n$ & & $\Delta d$ & $\Delta n$ \\ \hline \hline
0.5 & 6 & 66 & 132 & 6.6 & 11.8 &  & -2 (-3\%) & 4 (+3\%) &  & -2 (-3\%) & 11 (+8\%) \\ 
   &  & 66 & 110 & 5.5 & 14.1 &  & -2 (-3\%) & 2 (+2\%) &  & -1 (-2\%) & 8 (+7\%) \\ 
   &  & 66 & 96 & 4.8 & 17.0 &  & -2 (-3\%) & 1 (+1\%) &  & 0 (+0\%) & 7 (+7\%) \\ 
   &  & 66 & 84 & 4.2 & 22.0 &  & -1 (-2\%) & 2 (+2\%) &  & 3 (+5\%) & 9 (+11\%) \\ 
   & 12 & 66 & 132 & 6.6 & 20.2 &  & -2 (-3\%) & 3 (+2\%) &  & -2 (-3\%) & 8 (+6\%) \\ 
   &  & 66 & 110 & 5.5 & 25.4 &  & -2 (-3\%) & 2 (+2\%) &  & -1 (-2\%) & 7 (+6\%) \\ 
   &  & 66 & 96 & 4.8 & 31.8 &  & -2 (-3\%) & 1 (+1\%) &  & 1 (+2\%) & 8 (+8\%)) \\ 
   &  & 66 & 84 & 4.2 & 42.4 &  & 0 (+0\%) & 3 (+4\%) &  & 3 (+5\%) & 9 (+11\%) \\ 
   & 24 & 66 & 132 & 6.6 & 37.3 &  & -2 (-3\%) & 3 (+2\%) &  & -2 (-3\%) & 8 (+6\%) \\ 
   &  & 66 & 110 & 5.5 & 48.7 &  & -2 (-3\%) & 2 (+2\%) &  & -1 (-2\%) & 7 (+6\%) \\ 
   &  & 66 & 96 & 4.8 & 62.4 &  & -2 (-3\%) & 1 (+1\%) &  & 1 (+2\%) & 8 (+8\%) \\ 
   &  & 66 & 84 & 4.2 & 85.5 &  & 0 (+0\%) & 3 (+4\%) &  & 4 (+6\%) & 10 (+12\%) \\ 
  0.6 & 6 & 121 & 242 & 8.1 & 12.0 &  & -2 (-2\%) & 8 (+3\%) &  & 2 (+2\%) & 37 (+15\%) \\ 
   &  & 121 & 202 & 6.7 & 13.8 &  & -2 (-2\%) & 5 (+2\%) &  & 3 (+2\%) & 22 (+11\%) \\ 
   &  & 121 & 174 & 5.8 & 16.5 &  & -1 (-1\%) & 4 (+2\%) &  & 5 (+4\%) & 18 (+10\%) \\ 
   &  & 121 & 152 & 5.1 & 21.0 &  & 0 (+0\%) & 4 (+3\%) &  & 8 (+7\%) & 17 (+11\%) \\ 
   & 12 & 121 & 242 & 8.1 & 19.6 &  & -2 (-2\%) & 6 (+2\%) &  & 1 (+1\%) & 22 (+9\%) \\ 
   &  & 121 & 202 & 6.7 & 24.1 &  & -2 (-2\%) & 4 (+2\%) &  & 3 (+2\%) & 19 (+9\%) \\ 
   &  & 121 & 174 & 5.8 & 30.2 &  & -1 (-1\%) & 4 (+2\%) &  & 5 (+4\%) & 17 (+10\%) \\ 
   &  & 121 & 152 & 5.1 & 39.9 &  & 0 (+0\%) & 4 (+3\%) &  & 9 (+7\%) & 18 (+12\%) \\ 
   & 24 & 121 & 242 & 8.1 & 35.3 &  & -2 (-2\%) & 6 (+2\%) &  & 1 (+1\%) & 19 (+8\%) \\ 
   &  & 121 & 202 & 6.7 & 45.2 &  & -2 (-2\%) & 4 (+2\%) &  & 3 (+2\%) & 18 (+9\%) \\ 
   &  & 121 & 174 & 5.8 & 58.4 &  & -1 (-1\%) & 4 (+2\%) &  & 5 (+4\%) & 16 (+9\%) \\ 
   &  & 121 & 152 & 5.1 & 79.6 &  & 0 (+0\%) & 4 (+3\%) &  & 9 (+7\%) & 18 (+12\%) \\ 
  0.7 & 6 & 247 & 494 & 12.4 & 14.0 &  & NA & NA &  & NA & NA \\ 
   &  & 247 & 412 & 10.3 & 15.1 &  & 1 (+0\%) & 18 (+4\%) &  & 14 (+6\%) & 77 (+19\%) \\ 
   &  & 247 & 354 & 8.9 & 17.3 &  & 2 (+1\%) & 13 (+4\%) &  & 17 (+7\%) & 48 (+14\%) \\ 
   &  & 247 & 310 & 7.8 & 21.0 &  & 4 (+2\%) & 11 (+4\%) &  & 23 (+9\%) & 43 (+14\%) \\ 
   & 12 & 247 & 494 & 12.4 & 20.9 &  & 0 (+0\%) & 19 (+4\%) &  & 11 (+4\%) & 72 (+15\%) \\ 
   &  & 247 & 412 & 10.3 & 24.5 &  & 1 (+0\%) & 14 (+3\%) &  & 14 (+6\%) & 50 (+12\%) \\ 
   &  & 247 & 354 & 8.9 & 29.8 &  & 2 (+1\%) & 11 (+3\%) &  & 17 (+7\%) & 41 (+12\%) \\ 
   &  & 247 & 310 & 7.8 & 38.2 &  & 4 (+2\%) & 11 (+4\%) &  & 23 (+9\%) & 40 (+13\%) \\ 
   & 24 & 247 & 494 & 12.4 & 35.3 &  & 0 (+0\%) & 15 (+3\%) &  & 11 (+4\%) & 54 (+11\%) \\ 
   &  & 247 & 412 & 10.3 & 44.0 &  & 1 (+0\%) & 12 (+3\%) &  & 13 (+5\%) & 42 (+10\%) \\ 
   &  & 247 & 354 & 8.9 & 55.9 &  & 2 (+1\%) & 10 (+3\%) &  & 17 (+7\%) & 38 (+11\%) \\ 
   &  & 247 & 310 & 7.8 & 74.3 &  & 4 (+2\%) & 10 (+3\%) &  & 24 (+10\%) & 40 (+13\%) \\ 
  0.8 & 6 & 631 & 1262 & 25.2 & 21.3 &  & NA & NA &  & NA & NA \\ 
   &  & 631 & 1052 & 21.0 & 21.3 &  & NA & NA &  & NA & NA \\ 
   &  & 631 & 902 & 18.0 & 22.1 &  & 14 (+2\%) & 72 (+8\%) &  & NA & NA \\ 
   &  & 631 & 790 & 15.8 & 24.7 &  & 17 (+3\%) & 39 (+5\%) &  & 66 (+10\%) & 144 (+18\%) \\ 
   & 12 & 631 & 1262 & 25.2 & 27.5 &  & NA & NA &  & NA & NA \\ 
   &  & 631 & 1052 & 21.0 & 29.4 &  & 11 (+2\%) & 60 (+6\%) &  & 52 (+8\%) & 272 (+26\%) \\ 
   &  & 631 & 902 & 18.0 & 33.4 &  & 13 (+2\%) & 39 (+4\%) &  & 57 (+9\%) & 138 (+15\%) \\ 
   &  & 631 & 790 & 15.8 & 40.4 &  & 16 (+3\%) & 31 (+4\%) &  & 66 (+10\%) & 110 (+14\%) \\ 
   & 24 & 631 & 1262 & 25.2 & 40.5 &  & 8 (+1\%) & 57 (+5\%) &  & 47 (+7\%) & 229 (+18\%) \\ 
   &  & 631 & 1052 & 21.0 & 47.2 &  & 10 (+2\%) & 39 (+4\%) &  & 51 (+8\%) & 143 (+14\%) \\ 
   &  & 631 & 902 & 18.0 & 57.5 &  & 12 (+2\%) & 31 (+3\%) &  & 57 (+9\%) & 114 (+13\%) \\ 
   &  & 631 & 790 & 15.8 & 73.7 &  & 16 (+3\%) & 29 (+4\%) &  & 67 (+11\%) & 103 (+13\%) \\
\hline
\end{tabular}
\end{table}